\documentclass[twocolumn,aps,prl,showpacs,amsmath,superscriptaddress]{revtex4}
\usepackage{amsmath,amssymb,mathrsfs}

\usepackage{hyperref}
\usepackage{paralist}
\usepackage{pdfpages}
\usepackage{color}
 \usepackage{graphicx}

\def\be{\begin{equation}}
\def\ee{\end{equation}}
\def\bea{\begin{eqnarray}}
\def\eea{\end{eqnarray}}

\begin{document}

\title{New trends in quantum integrability:   Recent  experiments with ultracold atoms}

\author{Xi-Wen Guan}
\email[e-mail:]{xwe105@wipm.ac.cn; xiwen.guan@anu.edu.au}
\affiliation{State Key Laboratory of Magnetic Resonance and Atomic and Molecular Physics,
Wuhan Institute of Physics and Mathematics, APM, Chinese Academy of Sciences, Wuhan 430071, China}
\affiliation{NSFC-SPTP Peng Huanwu Center for Fundamental Theory, Xi'an 710127, China}
\affiliation{Department of Fundamental and Theoretical Physics, Research School of Physics,
Australian National University, Canberra ACT 0200, Australia}

\author{Peng He}
\email[e-mail:]{phe@cashq.ac.cn}
\affiliation{Bureau of Frontier Sciences and Education, Chinese Academy of Sciences, Beijing 100864, China
}

\begin{abstract}

Over the past two decades quantum engineering has made significant advances in our ability to create genuine quantum many-body systems using ultracold atoms.
In particular, some prototypical exactly solvable Yang-Baxter  systems have been successfully realized  allowing  us to confront elegant and
sophisticated exact solutions  of these systems with their experimental counterparts.
The new experimental developments    show a variety of fundamental one-dimensional (1D) phenomena,
 ranging from the generalized hydrodynamics to  dynamical fermionization, Tomonaga-Luttinger liquids, collective excitations, fractional exclusion
 statistics, quantum holonomy, spin-charge separation, competing orders with  high spin symmetry and  quantum impurity problems.
This article briefly reviews  these developments  and provides rigorous  understanding of  those observed phenomena  based on the exact solutions while highlighting the uniqueness of 1D quantum  physics.
 The precision of atomic physics   realizations of  integrable many-body problems continues to inspire significant developments in mathematics and physics while at the same time offering   the prospect  to contribute to future quantum technology.

\end{abstract}
\author{}\maketitle

\section{ I. Introduction}

Exactly solvable models  date back to 1931 when Bethe \cite{Bethe:1931} introduced a particular form of wave function to obtain eigenspectrum of one-dimensional (1D) Heisenberg spin chain.
His seminal many-body wave function which consisted of a superposition of all possible plane waves was later called Bethe ansatz.
However, it was more than 30 years after his work that the Bethe ansatz  was further applied to other physics problems in 1D, including the Lieb-Liniger Bose gas \cite{Lieb-Liniger:1963}, the Yang-Gaudin Fermi gas \cite{Yang:1967,Gaudin:1967}, the 1D Hubbard model \cite{Lieb:1968},  the SU(N) Fermi gases \cite{Sutherland:1968}, the Kondo impurity problems \cite{ Andrei:1980, Wiegmann:1980,Tsvelik:1983,Andrei:1983} and the Bardeen-Cooper-Schrieffer (BCS) pairing model \cite{BCS:1957, Richardson:1963,Dukelsky:2004} among others.
For a long time these exactly solvable models were considered to be mathematical toy models despite they had inspired significant developments in both mathematics and physics.

\begin{figure}[th]
 \begin{center}
 \includegraphics[width=0.9\linewidth]{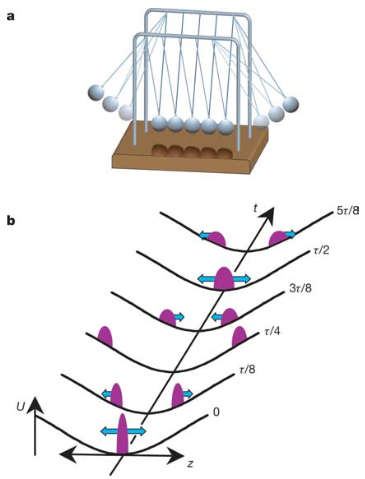}
 \end{center}
 \caption{ A demonstration of classical  Newton's cradle (a)  and quantum  Newton's cradle (b). In the later,  initially separated two parts of  the cloud of quasi-1D trapped Bose gas  in momentum space oscillate out-of-equilibrium for several dozens  of ms,  showing  a  quantum version of Newton's cradle. This novel observation of time evolution of the integrable model out-of-equilibrium provided impetus for the development of generalized hydrodynamics \cite{Castro:2016,Bertini:2016}. Figure from \cite{Kinoshita:2006}. }
 \label{fig:Quantum-cradle}
\end{figure}

However, this impression was changed dramatically  since the realization of one-dimensional integrable models of ultracold atoms in labs in the last two decades.  Consequently, the study of exactly solvable models  saw  a renewed  interest,  see for instance  the reviews \cite{Cazalilla:2011,Guan:2013,Batchelor:2016,Mistakidis:2022}.

This breakthrough was a result of a series of spectacular advances in the experimental control and manipulation of ultracold atoms, involving the trapping and cooling of atomic particles to extremely low temperatures of few nano-Kelvin above the absolute zero \cite{Giorgini:2008,Lewenstein:2007}.
It led to finding new phases of cold matter resulting  from the effects of interactions,  statistics and symmetries of many atoms which requires deep understanding  of coherent and correlation  nature of many interacting particles at a higher level of rigor.

The experimental realizations of 1D Bose and Fermi gases with tunable interaction and internal degrees of freedom between atoms have provided a remarkable testing ground for quantum integrable systems.

Strikingly different features of many-body effects result from quantum degeneracy, i.e. bosons with integer spin obey Bose-Einstein statistics, whereas fermions with half odd integer spin obey Fermi-Dirac statistics.
However, interactions among such degenerate cold particles can dramatically change their behaviour, resulting in novel many-body phenomena, for example, superfluid, such as Bose-Einstein condensation (BEC), quantum liquid, criticality etc.

The observed results to date are seen to be in excellent agreement with predictions from  the  exactly solved models.
Early on, the prototypical integrable  model--the Lieb-Liniger Bose gas was realized to reveal ground state properties, local pair correlation \cite{Kinoshita:2004,Paredes:2004,Kinoshita:2005}.
Subsequently, the Yang-Yang thermodynamics  and quantum fluctuation in  the model were  observed \cite{Amerongen:2008,Armijo:2010,Jacqmin:2011,Stimming:2010,Armijo:2012,Vogler:2013}.
Based on theoretical prediction of the fermionization in the Lieb-Liniger gas \cite{Astrakharchik:2005,Batchelor:2005}, the novel super Tonks-Girardeau gas-like state  was  experimentally realized  in \cite{Haller:2009,Kao:2021}.
Observation of the quantum degenerate spin-1/2 Fermi gas \cite{Moritz:2005,Liao:2010,Zurn:2012,Zurn:2013,Zurn:2013b,Wenz:2013,Murmann:2015,Revelle:2016} and SU(N) Fermi gases \cite{Pagano:2014,Song:2020} are having high impact in ultracold atoms \cite{Wu:2003,Wu:2006}.

New trends in experiments with Yang-Baxter quantum  integrability involve quantum criticality and Luttinger liquid \cite{Haller:2010,Fabbri:2015,Meinert:2015,Yang:2017,YangTL:2018}, quantum dynamics, thermalization, correlation \cite{Schweigler:2017,Prufer:2017,Wilson:2020,Sun:2021,Tang:2018,Erne:2018}.
Very recently, the 1D Hubbard model  \cite{Boll:2016,Hilker:2017,Vijayan:2020,Spar:2022}, 
 1D p-wave  polarized fermions \cite{Chang:2020,Ahmed-Braun:2021,Jackson:2022,Venu:2022}, 1D matter wave breathers \cite{Yurovsk:2017,Luo:2020,Marchukov:2020} and emergent dynamical quasicondensation of hard-core bosons at finite momenta \cite{Vidmar:2015}  also found their  ways into the  lab.
As a result,  the mathematical physics of Bethe ansatz integrable models has not only become testable in experiments with ultracold atoms but  the experiments also provided impetus for many new developments of quantum integrability.
For example, the realization of the quantum Newton cradle in the 1D Lieb-Liniger gas \cite{Kinoshita:2006} led to a  new surge of study of integrable models.
In particular, the experimental observation of the quantum Newton's cradle  \cite{Kinoshita:2006} raised a significant question for theory and experiment: can the evolution of isolated quantum integrable systems of many particles  converge into its equilibrium Gibbs thermal state?
Seeking for an understanding of such a subtlety  of quantum dynamics of quantum integrable models  leads to new developments of the Generalized Gibbs ensemble (GGE) \cite{Rigol:2007,Ilievski:2015} and
generalized hydrodynamics (GHD) \cite{Castro:2016,Bertini:2016}, which  have  become a promising theme in theory and experiment with 1D ultracold atoms \cite{Langen:2015,Schemmer:2019,Malvania:2021,Berg:2016,Caux:2019,Nardis:2018,Bastianello:2019,Bastianello:2020,Ruggierro:2020,Doyon:2017,Fava:2021,Bastianello:2022,Bouchoule:2022,Moller:2022,Bertini:2022,Moller:2021,Li-Chen:2020,Cataldini:2021}.

Last but not least, the quantum walks of one magnon and  magnon bound states in interacting bosons on 1D lattices \cite{Fukuhara:2013a,Fukuhara:2013b}, realization of 1D antiferromagnetic Heisenberg spin chain with ultracold spinor Bose gas \cite{Sun:2021},
quantum impurities \cite{Catani:2012,Meinert:2017}, fractional quantum statistics  \cite{Zhang:2022}  and spin-charge separation \cite{Senaratne:2021}
open new frontiers in  quantum technology.

\section{II. Exactly solved models in ultracold atoms}

 The study of Bethe ansatz solvable models flourished in the period from 1960s to 1990s in the  mathematical physics communities in  the world.
A number of notable Bethe ansatz integrable models in a variety of fields of physics were solved at that time.
In the last two decades, a considerable attention  has been paid to the study of  1D exactly solved models due to unprecedented abilities  in cooling and controlling cold atoms in low dimensions.
The exact results of the integrable models provide deep insight into the quantum simulation and understanding of many known as yet unknown quantum phases of matter in ultracold atoms, condensed matter physics and quantum metrology.
Here we first introduce several exactly solvable models which have provided rigorous understanding of fundamental many-body physics observed so far in low-dimensional ultracold atoms and will give further possibilities for exploring subtle quantum phenomena in 1D.

\subsection{Lieb-Liniger Bose gas}
 \label{subsection-LL-gas}
The Hamiltonian of the  Lieb-Liniger  gas  of
$N$ particles  with periodic boundary conditions in a  length $L$  is given by   \cite{Lieb-Liniger:1963}
\begin{eqnarray}
  \hat{H} =
  \frac{\hbar^2}{2 m}\int_0^L {\rm d} x~\partial_x\hat{\psi}^\dag
  \partial_x \hat\psi
  + \frac{g_{\rm 1D}}{2} \int_0^L {\rm d} x ~\hat\psi^\dag
  \hat\psi^\dag\hat\psi\hat\psi, \label{Ham}
\end{eqnarray}
where  $m$ is the mass of the bosons, $g_{\rm 1D}$ is  the coupling constant which is determined  by the the 1D scattering length  $g_{\rm 1D} = -2 \hbar^2/m a_{\rm 1D}$.
The scattering length  is given by $a_{1D}=\left( -a_{\perp }^{2}/2a_{s}\right) \left[ 1-C\left(
a_{s}/a_{\perp }\right) \right] $ \cite{Olshanii_PRL_1998,Dunjko_PRL_2001,Olshanii_PRL_2003}.
In the above equation, the canonical quantum Bose fields $\hat\psi(x)$   satisfying the following  commutation relations
\begin{eqnarray}
\begin{split}
 &\big[\hat\psi(x),\hat\psi^\dag(y)\big]=\delta(x-y),\\
 &\big[\hat\psi(x),\hat\psi(y)\big]=
 \big[\hat\psi^\dag(x),\hat\psi^\dag(y)\big]=0.
\end{split}
\nonumber
\end{eqnarray}
Solving the Schr\"odinger equation
${H}\varPsi(x)=E\varPsi(x)$ reduces to the eigenvalue problem of
the many-particles Hamiltonian
\begin{eqnarray}
  H = - \sum_{i = 1}^N \frac{\partial^2}{\partial x_i^2} + 2 c \sum_{i < j}^N
  \delta ( x_i - x_j).\label{Ham1}
\end{eqnarray}
Here the effective coupling constant $c=\frac{g_{\rm 1D}}{2}$.

Following the Bethe ansatz \cite{Bethe:1931} the wave function is divided into $N!$ domains  according to the positions of the particles $\varTheta({\cal Q}): x_{{\cal Q}_1}< x_{{\cal Q}_2}<\cdots< x_{{\cal Q}_N}$, where $\cal Q$ is the permutation of number set $\{1,2\cdots, N\}$.
The wave function can be written as $\varPsi(\boldsymbol{x})=\sum_{\cal Q}\varTheta({\cal Q}) \psi_{\cal Q}(\boldsymbol{x})$.
Considering the symmetry  of bosonic statistics,  all the $\psi$ in different domain  ${\cal Q}$ should be  the same, i.e., $\psi_{\cal Q}=\psi_{{\cal I}}$, where we use ${{\cal I}}$ to denote the elements of the identity permutation, ${{\cal I}}=\{1,2,\cdots,N\}$.
Lieb and Liniger   \cite{Lieb:1968} wrote the wave function for  the model (\ref{Ham1})  as the superposition of $N!$ plane waves
\begin{eqnarray}
 \label{WF-Lieb}
 \psi_{{\cal I}} (x) =\sum_{\cal P}A({\cal P}){\rm e}^{{\mathrm i}(k_{{\cal P}_1}x_{1}+\cdots+k_{{\cal P}_N}x_{N})},
\end{eqnarray}
  where ${\cal P}$ denotes  the permutation of number set $\{1,2\cdots, N\}$ with  the pseudo-momenta  $k_i$s  carried by the particles under   periodic boundary conditions.
They satisfy  a set
of Bethe ansatz equations, called the Lieb--Liniger equations
\begin{equation}
  \mathrm{e}^{\text{i} k_j L} = - \prod_{l = 1}^N \frac{k_j - k_l + \text{i} c}{k_j - k_l,
  - \text{i} c}.\label{BAE}
\end{equation}
Here $j=1,\ldots, N$.
For a given set of quasi-momenta
$\left\{ k_j\right\}$, the total momentum and the energy of the
system are obtained by
\begin{eqnarray}
 P=\sum_{j}^N k_j,\qquad
 E=\sum_{j}^N k^2_j.
\end{eqnarray}
The fundamental physics of the model  (\ref{Ham1}) are essentially determined by the wave function (\ref{WF-Lieb}) and the Lieb--Liniger equations (\ref{BAE}), see a review \cite{Jiang:2015}.

 In 1969  C N Yang and C P Yang  \cite{Yang-Yang:1969} published their  seminal work on  the thermodynamics of the Lieb-Liniger Bose gas.
 The thermodynamics of the Lieb-Liniger gas is  determined from the minimization conditions of the
Gibbs free energy  in terms of the  roots of  the Bethe ansatz equations (\ref{BAE}).
In the thermodynamic limit,  the Bethe ansatz equations (\ref{BAE}) become
\begin{equation}
\rho(k)+\rho^{h}(k)=\frac{1}{2\pi}+\frac{c}{\pi}\int_{-\infty}^{\infty}\frac{\rho(k')dk'}{c^{2}+(k-k')^{2}},\label{BA-PH}
\end{equation}
where $\rho(k)$ and   $\rho^{h}(k)$ are respectively the particle  and hole density distribution functions at finite temperatures.
The Gibbs free energy per unit length is
given by $G/L=E/L-\mu n-Ts$ with the relation to the free energy
 $F=G+\mu N$. Here $\mu$ is the chemical potential, $n$ is the linear  density, and $s$ is  the entropy.
 The minimization condition  $\frac{\delta G}{L}=0$ with respect to particle density $\rho$  leads to
the  Yang-Yang thermodynamic Bethe ansatz  (TBA)   equation  \cite{Yang-Yang:1969}
\begin{equation}
\varepsilon(k)=k^{2}-\mu
-\frac{Tc}{\pi}\int_{-\infty}^{\infty}\frac{dq}{c^{2}+(k-q)^{2}}\ln\left(1+e^{-\frac{\varepsilon(q)}{T}}\right), \label{TBA-Bose}
\end{equation}
which determines the thermodynamics of the system in the whole temperature regime through
the Gibbs free energy per length, i.e.,
\begin{eqnarray}
p &=&  -\left(\frac{\partial G}{\partial L}\right)_{T,\mu,c}=\frac{T}{2\pi}\int\ln\left(1+e^{-\varepsilon(k)/T}\right)dk.\label{pressure}
\end{eqnarray}
Thus  other thermodynamic quantities can be obtained
through  thermodynamic relations, for example
the particle density, entropy
density, compressibility, specific heat are given by
\begin{eqnarray}
n& =& \partial_\mu p|_{c, T},\qquad s=\partial_T p|_{\mu, c}\nonumber \\
\kappa&=&\partial^2_\mu p|_{c, T},\qquad c_{\rm
v}=T\partial^2_T p|_{\mu,c}.
\end{eqnarray}
In the above equation $\partial_x $ denotes the derivative respect to the potentials.
Exact Bethe ansatz solution provides an analytical way to access to  quantum critical behaviour  of the system at finite  temperatures \cite{Guan:2011,Jiang:2015}.

\subsection{Yang-Gaudin model}

1D models of Bethe ansatz solvable quantum gases   have  become an important  subject of experiment, such as  two-component spinor  fermions and bosons  as well as multi-component fermions.
Models of this kind have rich internal degrees of freedom and high symmetries.
In 1967, C N Yang \cite{Yang:1967} solved the 1D delta-function interacting Fermi gas wit the discovery of quantum integrability,
i.e. the necessary condition for the Bethe ansatz solvability.
This quantum integrability condition was referred to the factorization condition--the scattering matrix of a quantum many-body system can be factorized into a product of many two-body scattering matrices.
At the same time, M Gaudin also rigorously derived the Bethe ansatz  equations for the spin-$1/2$ Fermi gas for spin balance case \cite{Gaudin:1967}.
Therefore, the 1D spin-$1/2$ Fermi gas  is now called the Yang-Gaudin model.
On the other hand, in 1972,  R J   Baxter \cite{Baxter:1972}  independently found that a similar  factorization relation also occurred as the conditions for commuting transfer matrices in 2D  vertex  models in statistical mechanics.
Such a factorization condition is  now known as the Yang-Baxter equation and becomes a significant  key condition  to quantum integrability.

The many-body Hamiltonian for the Yang-Gaudin  model is given by   \cite{Yang:1967,Gaudin:1967}
\begin{equation}
H=-\frac{\hbar^{2}}{2m}\sum_{i=1}^{N}\frac{\partial^{2}}{\partial
x_{i}^{2}}+g_{1D}\sum^{'}_{1\leq i<j\leq
N}\delta(x_{i}-x_{j}) \label{Ham2}
\end{equation}
 describes $N$  fermions of the same mass $m$ with two internal spin  states  confined to a  length $L$ interacting via a $\delta$-function potential.  In the above equation, prime denotes the exclusion of the interaction between two atoms with the same spin. 
Here we denoted the numbers of fermions in  the two  hyperfine levels $|\uparrow \rangle $ and $|\downarrow \rangle$  as  $N_{\uparrow}$ and $N_{\downarrow}$, respectively.
The total number of fermions and the magnetization were denoted as $N=N_{\uparrow}+ N_{\downarrow}$ and $M^z=(N_{\uparrow}-N_{\downarrow})/2$, respectively.
The  interaction strength is given by $g_{1D}=\hbar^{2}c/m$ with  $c=-2/a_{1D}$, where
$a_{1D}$ is the effective 1D scattering length, see the  discussion on the Hamiltonian of the Lieb-Liniger gas. 
For convenience, we also  define a dimensionless interaction strength $\gamma = c / n$. 
Here the linear density  is defined  by $n = N / L$; meanwhile, $c>0$ for   a repulsive interaction and  $c<0$ for an attractive interaction, see review \cite{Guan:2013}. 

C N Yang generalized Bethe's wave function to the following  many-body wave function of the spin-$1/2$ Fermi gas
\begin{equation}
\psi=\sum_{P}A_{\sigma_{1}\ldots\sigma_{N}}(P|Q)\exp
\textrm{i}(k_{P1}x_{Q1}+\ldots+k_{PN}x_{QN})
\label{wavefunction}
\end{equation}
for  the domain
$0<x_{Q1}<x_{Q2}<\ldots<x_{QN}<L$.
Where $\{k_{i}\}$  denote  a set of unequal wave  numbers and  $\sigma_{i}$ with $i=1,\ldots, N$   indicate the spin coordinates. Both $P$ and $Q$
are permutations of  indices $\{1,2,\ldots,N\}$, i.e.  $P=\left\{ P_{1},\ldots,P_{N}\right\}$ and $Q=\left\{ Q_{1},\ldots,Q_{N}\right\}$.  The sum runs over all $N!$ permutations $P$ and the coefficients of the exponentials are column
vectors with each of the $N!$ components representing a permutation
$Q$.
For determining  the wave function associated with the irreducible representations of the permutation group $S_N$ and the irreducible representation of the Young tableau for different  up- and down-spin fermions,
C N Yang found a key condition for solvability, i.e.,
the two-body scattering matrix acting on three  linear tensor spaces $V_1\otimes V_2\otimes V_3$
\begin{equation}
Y_{ij}(u)=\frac{\textrm{i}uT_{ij}+cI}{\textrm{i}u-c}
\end{equation}
satisfies the following cubic equation
\begin{eqnarray}
&&Y_{12}(k_{2}-k_{1})Y_{23}(k_{3}-k_{1})Y_{12}(k_{3}-k_{2})\nonumber \\
&& =Y_{23}(k_{3}-k_{2})Y_{12}(k_{3}-k_{1})Y_{23}(k_{2}-k_{1}),\label{YBE}
\end{eqnarray}
which has been known as  the Yang-Baxter equation \cite{Yang:1967,Baxter:1972}.
It is worth mentioning that an experimental simulation of the Yang-Baxter equation itself through the linear quantum optics \cite{Zhang:2013}
and the  Nuclear Magnetic Resonance interferometric setup \cite{Vind:2016} has been performed.
In the above equations,  we denoted the operator $T_{ij}=-P_{ij}$, here
$P_{ij}$ is the permutation operator.
The Yang-Baxter equation has found a profound legacy in both mathematics and physics, see review \cite{1D-Hubbard, Korepin,Sutherland-book,Takahashi-b,Wang-book,Guan:2013}.

Solving the eigenvalue
problem of $N$ interacting  particles  (\ref{Ham2}) in a periodic box of length $L$ with  the periodic boundary conditions,
C N Yang obtained the  Bethe ansatz equations for the  Fermi gas  \cite{Yang:1967}
\begin{eqnarray}
&&{\rm e}^{i k_j L} =\prod_{\alpha =1}^{M} \frac{k_j -\lambda_{\alpha} +ic/2}{k_j -\lambda_{\alpha} -ic/2}, \label{BA1} \\
&&\prod_{j=1}^N  \frac{\lambda_{\alpha}-k_j +ic/2}{\lambda_{\alpha}-k_j -ic/2}=-\prod_{\beta=1}^{M}\frac{\lambda_{\alpha}-\lambda_{\beta}+ic}{\lambda_{\alpha}-\lambda_{\beta}-ic},\label{BA2}
\end{eqnarray}
with $j=1,2,\cdots,N$ and $\alpha=1,2,\cdots,M$. Here $M$ is the number of atoms with down-spins.
The energy eigenspectrum is given in terms of the quasimomenta $\left\{k_i\right\}$  of the fermions via
$E=\frac{\hbar ^2}{2m}\sum_{j=1}^Nk_j^2$.
In the above equations (\ref{BA1}) and (\ref{BA2}), $\lambda_{\alpha} $ denoted spin rapidities.
All quasimomenta $\left\{k_i\right\}$  are distinct and uniquely determine the wave function of model Eq.(\ref{wavefunction}).

The fundamental physics of the model (\ref{Ham2})  can be   obtained by solving the transcendental Bethe ansatz equations (\ref{BA1}) and (\ref{BA2}).
Finding the solution of the Bethe ansatz equations (\ref{BA1}) and (\ref{BA2}) is extremely  cumbersome.
For a repulsive interaction,  in
the thermodynamic limit, i.e., $L,N \to \infty$, and $N/L$ is finite, 
 the above Bethe ansatz equations   can be written as the generalized Fredholm equations
\begin{eqnarray}
{\rho}(k)&=&\frac{1}{2\pi}+ \int_{-B_2}^{B_2}a_1(k-\lambda){\sigma_1 }(\lambda )d\lambda, \label{BE2-r1}\\
{\sigma_1}(\lambda )&=&\int_{-B_1}^{B_1}a_1(\lambda -k){\rho }(k)dk\nonumber \\
&& - \int_{-B_2}^{B_2}a_2(\lambda-\lambda'){\sigma_1}(\lambda') d\lambda'.\label{BE2-r2}
\end{eqnarray}
The associated   integration boundaries $B_1$, $B_2$   are determined by the relations
\begin{eqnarray}
n_{t}:&\equiv &N/L=\int_{-B_1}^{B_1}{\rho}(k)dk,  \nonumber\\
 n_{\downarrow}:&\equiv& N_{\downarrow}/L=\int_{-B_2}^{B_2}{\sigma_1}(k)dk,\label{repulsive-d}
\end{eqnarray}
where  $n_{t}$ denotes  the linear density, while $n_{\downarrow}$ is the density of spin-down fermions.
In the above equations we introduced the quasimomentum distribution function $\rho(k)$ and distribution function of the spin rapidity $\sigma_1(\lambda)$ for the ground state.

The thermodynamics of the Yang-Gaudin model with a repulsive interaction was studied by C K  Lai \cite{Lai:1971,Lai:1973} and  M Takahashi \cite{Takahashi:1971}.
In the same fashion as the derivation of the thermodynamics of the Lieb-Liniger Bose gas by C. N. Yang and C. P. Yang in 1969 \cite{Yang-Yang:1969},
 M Takahashi \cite{Takahashi:1971} gave the  thermodynamic Bethe ansatz equations for the repulsive Yang-Gaudin model
\begin{eqnarray}
	\varepsilon (k) &=& k^2-\mu-\frac{H}{2}-T \sum_{n=1 }^{\infty}a_n*{\rm ln} [1+{\rm e}^{- \phi_n (\lambda)/T}], \label{wholeTBA1}\quad \\
	\phi_n (\lambda)&=& nH-T a_n*\ln [1+{\rm e}^{- \varepsilon (k)/T}]\nonumber \\
	&+& T\sum_{m=1}^{\infty} T_{mn}*{\rm ln} [1+{\rm e}^{- \phi_m (\lambda)/T}] \label{wholeTBA2}
\end{eqnarray}
with $n=1,\ldots, \infty$.
Here $*$ denotes the convolution,  i.e. $\hat a\ast f(x)=\int_{-\infty}^{\infty}
a(x-y)f(y){\rm d}y$,  $\varepsilon(k)$ and $\phi_n(\lambda)$ are the dressed energies for the charge and the length-$n$ spin strings, respectively, with $k$'s and $\lambda$'s being the rapidities; and the function $T_{mn}(x)$ is  given by
\begin{eqnarray}
T_{mn}(\lambda)=
\left\{
\begin{array}{ll}
a_{|n-m|}(\lambda)+2a_{|n-m|+2}(\lambda)+\cdots \\
+ 2a_{m+n -2}(\lambda)+a_{m+n}(\lambda) &\text{for}\; m\ne n\\
2a_2 (\lambda) +2a_{4}(\lambda)+\cdots+a_{2n}(\lambda) &\text{for}\; m= n
\end{array}
\right..\nonumber
\end{eqnarray}
Where we denoted $ a_{m}(x) =\frac{1}{2\pi }\frac{mc}{(mc/2) ^{2}+x^{2}}$.
The pressure is given by
\begin{equation}
p=\frac{T}{2 \pi}\int_{-\infty}^{\infty} {\rm ln} [1+{\rm e}^{- \varepsilon (k)/T}] {\rm d}k,
\end{equation}
from which all the thermal and magnetic quantities, as well as other relevant physical properties, such as Wilson ratio \cite{Guan:2013b} and Gr\"uneisen parameter  \cite{Peng:2019,Yu:2020},   can be derived according to the standard statistical relations.

For a attractive interaction, i.e. $c<0$, the root patterns of the BA equations (\ref{BA1}) and (\ref{BA2}) are significantly different from that of the model for  $c>0$.
In this regime the quasimomenta $\left\{ k_i\right\} $ of the fermions with different spins  form two-body bound states \cite{Takahashi-a,Gu-Yang,Guan:2007},
 leading to the existence of the  Fulde-Ferrell-Larkin-Ovchinnikov (FFLO) pairing state \cite{Larkin1965,Fulde1964,Yang2001}. Here we will not discuss the attractive Yang-Gaudin model \cite{Orso:2007,Hu:2007,Guan:2007,Liu:2008,Zhao:2009,LeeGuan:2011,HeJS:2009}, a more detailed study of this model can be found in the review article \cite{Guan:2013}.

 \subsection{ Multi-component Fermi gases with $SU(w)$ symmetry}

It is shown that the fermionic alkaline-earth atoms display an exact $SU(w)$  spin symmetry with $w = 2I + 1$,  where $I$ is the nuclear spin \cite{Cazalilla:2009,Gorshkov:2010}.
 For example, the experiment \cite{Taie:2010} for ${}^{171}$Yb dramatically realised the model of fermionic atoms with $SU(2) \times SU(6)$ symmetry where electron spin decouples from its nuclear spin $I = 5/2$,
 and  $^{87}$Sr  atoms have $SU(10)$ symmetry \cite{XZhang2014S}.
 These experimental developments have opened   exciting opportunities to explore a wide range many-body phenomena such as spin and orbital magnetism \cite{Cappellini2014PRL, Scazza2014NP}, Kondo spin-exchange physics \cite{Nakagawa:2015,Zhang2015}  and  the one-dimensional (1D) Tomonaga-Luttinger liquid (TLL) \cite{Pagano:2014} etc.
 Such fermionic systems with enlarged $SU(w)$ spin symmetry are expected to display a remarkable diversity of new quantum phases and quantum critical phenomena due to the existence of multiple charge bound states \cite{Guan:2013,Wu:2006}.
 Here we do not wish to review the developments of $SU(w)$ Fermi gases in details.
We prefer to briefly  introduce the integrable  $SU(w)$ Fermi gases with repulsive interactions. 

We  introduce the Hamiltonian of $w$-component Fermi gas  of $N$ particles with mass $m$ \cite{Sutherland:1968}
\begin{eqnarray}
 \label{H}
 H=-\frac{\hbar ^{2}}{2m}\sum_{i=1}^{N}\frac{\partial^2}{\partial x_i^2} +g_{1D}\sum_{i<j}^{'}\delta(x_{i}-x_{j})+h \hat S_z,
\end{eqnarray}
where $\hat S_z$ is the total spin of the $z$-direction, $\hat S_z=
\sum_{r=1}^{w}[-(w+1)/2+r )]N_{r}$ and $h$ is the external magnetic field. In the above equation, the prime stands for the exclusion of the interaction between two atoms with the same spin.  Here $N_r$ is the particle number in the hyperfine  state $r$.
There are $w$ possible hyperfine states $|1\rangle, |2\rangle, \ldots,
|w\rangle$ that the fermions can occupy.
Again the interaction coupling constant is given by $g_\mathrm{1D} =-2\hbar^2/m a_\mathrm{1D}$, and  $c=m g_{1D}/\hbar^2$. Here  $a_\mathrm{1D}$ is the effective scattering length in 1D.
The Hamiltonian (\ref{H}) has the symmetry of $U(1)\times SU(w)$ when the magnetic field is absent, where $U(1)$ and $SU(w)$ are the symmetries of the charge and spin degrees of freedom, respectively.

Using Bethe's hypothesis, B Sutherland exactly solved the model by giving the  energy
 $E=\sum_{j=1}^Nk_j^2$,
and the Bethe ansatz equations
\begin{eqnarray}
 & {\rm e}^{\mathrm{i}k_{j}L}=\prod_{\ell=1}^{M_{1}}\frac
 {k_{j}-\lambda^{(w-1)}_{\ell}+{\rm i}\frac c2}
 {k_{j}-\lambda^{(w-1)}_{\ell}-{\rm i}\frac c2},
 j=1,\ldots,N. \label{BAE-SUN-1} \\
 &
 \prod_{j=1}^{N}\frac
 {\lambda^{(w-1)}_{\ell}-k_{j}+{\rm i}\frac c2}
 {\lambda^{(w-1)}_{\ell}-k_{j}-{\rm i}\frac c2}
 \prod_{m=1}^{M_{w-2}}\frac
 {\lambda^{(w-1)}_{\ell}-\lambda^{(w-2)}_{m}+{\rm i}\frac c2}
 {\lambda^{(w-1)}_{\ell}-\lambda^{(w-2)}_{m}-{\rm i}\frac c2}\nonumber\\
 &=-\prod_{\alpha =1}^{M_{w-1}}\frac
 {\lambda^{(w-1)}_{\ell}-\lambda^{(w-1)}_{\alpha}+\mathrm{i}c}
 {\lambda^{(w-1)}_{\ell}-\lambda^{(w-1)}_{\alpha}-\mathrm{i}c},
 \ell =1,\ldots ,M_{w-1}, \label{BAE-SUN-2}\\
 &
 \prod_{j=1}^{M_{r+1}} \frac
 {\lambda^{(r)}_{\ell}-\lambda^{(r+1)}_{j}+{\rm i}\frac c2}
 {\lambda^{(r)}_{\ell}-\lambda^{(r+1)}_{j}-{\rm i}\frac c2}
 \prod_{m=1}^{M_{r-1}}\frac
 {\lambda^{(r)}_{\ell}-\lambda^{(r-1)}_{m}+{\rm i}\frac c2}
 {\lambda^{(r)}_{\ell}-\lambda^{(r-1)}_{m}-{\rm i}\frac c2}\nonumber\\
 &=-\prod_{\alpha =1}^{M_{r}}\frac
 {\lambda^{(r)}_{\ell}-\lambda^{(r)}_{\alpha}+\mathrm{i}c}
 {\lambda^{(r)}_{\ell}-\lambda^{(r)}_{\alpha}-\mathrm{i}c},~~
\begin{array}{ll} r=2,3\cdots,w-2,\\
 \ell =1,2,\ldots ,M_{r},
 \end{array} \label{BAE-SUN-3} \\
 &
 \prod_{j=1}^{M_2} \frac
 {\lambda^{(1)}_{\ell}-\lambda^{(2)}_{j}+{\rm i}\frac c2}
 {\lambda^{(1)}_{\ell}-\lambda^{(2)}_{j}-{\rm i}\frac c2}
 =-\prod_{\alpha =1}^{M_1}\frac
 {\lambda^{(1)}_{\ell}-\lambda^{(1)}_{\alpha}+\mathrm{i}c}
 {\lambda^{(1)}_{\ell}-\lambda^{(1)}_{\alpha}-\mathrm{i}c}, \label{BAE-SUN-4}\\
 &
 \ell =1,\ldots ,M_1.\nonumber
\end{eqnarray}
Here $\{\lambda^{(r)}\}$, $r=1,2,\cdots,w-1$ denote the spin rapidities which are introduced  to describe the motion of spin waves. The particle number $N_r$ in each spin state links to the quantum number $M_\alpha$  via the relation $N_r=M_{r}-M_{r-1}$ and $M_0=0$.
For the repulsive interaction, i.e.  $c>0$,
there is no charge bound state, whereas each branch of spin rapidities $\{\lambda^{(r)}\}$ has complex roots
\begin{eqnarray}
 \lambda^{(r)}_{q,j,z} &= \lambda^{(r)}_{q,j}
 -\frac{\mathrm{i} c }{2}(q+1-2 z),\qquad z = 1,\cdots,q,
\end{eqnarray}
in  the thermodynamic limit, which are now called spin wave bound states.

Following Yang-Yang's grand canonical method \cite{Yang-Yang:1969}
 the thermodynamic Bethe ansatz  equations for  the Fermi gases (\ref{H}) are given by \cite{Takahashi-b,Guan:2013,Schlottmann:1997,Lee:2011,Jiang:2016}
\begin{eqnarray}
  \textstyle
 \varepsilon^{\rm c}(k)&= &k^{2}-\mu -sh
 +\sum_{n=1}^{\infty}\hat a_{n}\ast \varepsilon^{w-1,n}_-,\label{TBAE-SUN-1}
 \end{eqnarray}
 \begin{eqnarray}
  \textstyle
 \varepsilon^{w-1,n}(\lambda)&=&nh
 +\hat a_{n}\ast \varepsilon^{\rm c}_-
 +\sum_m \hat C_{n,m}\ast \varepsilon^{w-2,m}_-\nonumber\\
 &&
 -\sum_m \hat T_{mn}\ast \varepsilon^{w-1,m}_-,\label{TBAE-SUN-2}
 \end{eqnarray}
 \begin{eqnarray}
  \textstyle
 \varepsilon^{r,n}(\lambda)&=&nh
 +\sum_m \hat C_{n,m}\ast (\varepsilon^{r-1,m}_-+\varepsilon^{r+1,m}_-)\nonumber \\
 &&
 -\sum_m \hat T_{mn}\ast \varepsilon^{r,m}_-,\label{TBAE-SUN-3}
 \end{eqnarray}
 \begin{eqnarray}
 \textstyle
 \varepsilon^{1,n}(\lambda)&=&nh
 +\sum_m \hat C_{n,m}\ast \varepsilon^{2,m}_-
 -\sum_m \hat T_{mn}\ast \varepsilon^{1,m}_-.\label{TBAE-SUN-4}
\end{eqnarray}
Where $\varepsilon^{\rm c}(k)$ and $\varepsilon^{r,n}(k)$ are the dressed energies for  the charge sector and for  the branch $r$ in  the spin sector, respectively and  $\varepsilon_\pm= T\ln(1+{\rm e}^{\pm \varepsilon/T})$.  As a convention used in the TBA equations,  $n$ labels  the length of the strings, $\ast$ labels the convolution integral and the integral kernels are given by
\begin{eqnarray}
 T_{mn}&=& a_{m+n}+\ldots+2a_{|m-n|+2}+(1-\delta_{nm})a_{|m-n|},
 \nonumber\\
 C_{mn}&=&a_{m+n-1}+\ldots+a_{|m-n|+3}+a_{|m-n|+1}.
\end{eqnarray}
The pressure is given by
\begin{equation}
p=\frac{T}{2 \pi}\int_{-\infty}^{\infty} {\rm ln} [1+{\rm e}^{- \varepsilon (k)/T}] {\rm d}k,
\end{equation}
which serves as the equation of states for the repulsive $SU(w)$ Fermi gases.

The Bethe ansatz equations (\ref{BAE-SUN-1})-(\ref{BAE-SUN-4}) and the thermodynamic Bethe ansatz  equations  (\ref{TBAE-SUN-1})-(\ref{TBAE-SUN-4}) impose a big challenge in obtaining physical properties of the model \cite{Batchelor:2007,Schlottmann:1997,Lee:2011,Guan:2012a,Guan:2012b,Jiang:2016,Yang-You:2011}.
At low temperature, the behaviour of the Fermi gases with a repulsive interaction is described by spin-charge separated conformal field theories of an effective Tomonaga-Luttinger liquid and an antiferromagnetic $SU(w)$ Heisenberg spin chain, see experiment observation \cite{Pagano:2014}.
It was found \cite{Jiang:2016,Liu:2014,Schlottmann:1993} that the sound velocity of the Fermi gases in the large $w$ limit coincides with that for the spinless Bose gas, whereas the spin velocity for balance gases vanishes quickly as $w$ becomes large.
This novel feature of 1D multi-component Fermi gases shows a strong suppression of the Fermi exclusion statistics by the commutativity feature among the $w$-component fermions with different spin states in the Tomonaga-Luttinger liquid phase, which was for the first time proved by Yang and You  \cite{Yang-You:2011} in 2011, and confirmed by Guan and coworkers \cite{Guan:2012b} in 2012.
This feature was recently demonstrated with ultracold atoms in 2D \cite{Song:2020}.

Moreover, the 1D multi-component Fermi gases exhibit significantly different features from usual interacting electrons, showing rich pairing phenomena in momentum space, see review \cite{Guan:2013}.

\section{III. Generalized hydrodynamics of integrable systems}
Dynamical properties of isolated many-body systems have been the central problem in quantum statistics for a long time.
In the last decade, versatile dynamical properties of 1D integrable models have been demonstrated in the field of ultracold atoms.
In 2006, D. Weiss's group from the Pennsylvania State University   demonstrated nonequilibrium behaviour of interacting Bose gas in a quasi-1D  harmonic trap  in their  seminal paper \cite{Kinoshita:2006}, see Fig.~\ref{fig:Quantum-cradle}.
The clouds of quasi-1D trapped Bose gas  in momentum space oscillate out-of-equilibrium for several dozens  of ms that remarkably
illustrated behaviour of quantum Newton's cradle.
The framework to describe such kind of dynamics of integrable models out-of-equilibrium has been mainly  built on the GGE \cite{Rigol:2007,Ilievski:2015}
and the GHD  \cite{Castro:2016,Bertini:2016} with respect to an axial inhomogeneous trap \cite{Berg:2016,Caux:2019,Nardis:2018,Bastianello:2019,Bastianello:2020,Ruggierro:2020,Doyon:2017,Fava:2021}.
Further demonstrations of the GHD were carried out by using a quantum chip setup and the quasi-1D trap of ultracold atoms \cite{Schemmer:2019,Malvania:2021}.
So far, there have been a variety of theoretical methods for studying the Newton's cradle-like dynamics of 1D systems out-of-equilibrium.
Here we do not wish to review whole developments of the hydrodynamics. We will  introduce the basic  concepts of the GHD for later use in this paper.  More detailed theory  of the GHD can be seen in two nice  lecture notes by B Doyon \cite{Doyon:2017-L,Doyon:2020-L}.

 We  recapitulate  the GHD of the 1D integrable systems  introduced in the references  \cite{Castro:2016,Bertini:2016}.
 On the large scale, GHD is the classical  hydrodynamics, and for both we need a local equilibrium approximation, namely  a system displays a large scale motion which can be separated into many fluid  cells, and all these cells are large enough to be macroscopic and small enough to be homogeneous comparing to the size of  the system.
  The system can then be characterized by
	\begin{equation}
	\hat{\rho}=\hat{\rho}_{x_1}\otimes\hat{\rho}_{x_2}\otimes...\otimes\hat{\rho}_{x_n}\otimes...
	\end{equation}
	where $\hat{\rho}_{x_i}$ is the density matrix of an equilibrium state and $x_i$ is the position of the $i$th fluid cell.
In each cell, entropy maximisation occurs, i.e., reaching a local thermodynamic equilibrium.

On the other hand, for integrable systems with $N$ degrees of freedom, there are many local conserved quantities  ${\hat{Q}^1, \hat{Q}^2,...,\hat{Q}^N}, ...$, called conserved charges, the fundamental objects for quantum integrability.
 An GGE state is a maximal entropy state and its density matrix is constrained by all the conserved quantities
 by \cite{Rigol:2007,Ilievski:2015}
	\begin{eqnarray}
	\hat{\rho}=\frac{e^{-\sum_{i}\beta^i\hat{Q}^i}}{Tr[e^{-\sum_{i}\beta^i\hat{Q}^i}]}.
	\end{eqnarray}
Thus an equilibrium state of integrable systems can be identified by the expectation values of all the conserved quantities $Q^1, Q^2,...,Q^N,...$, where $Q^i=\langle \hat{Q}^i\rangle$, and $\beta^i$s  are the associated potentials.
For an non-equilibrium state, under local time-space equilibrium approximation,  integrable systems can be described by the distributions of all the conserved quantities  $q^1(x,t),q^2(x,t),...,q^N(x,t)$, where $x, t \in [0,L]$ and $q^i=\langle\hat{q}^i\rangle$,
	\begin{equation}
\hat{Q}^i=\int dx \hat{q}^i(x).
\end{equation}
Let us denote densities of conserved charges $\underline{q}\equiv (q^1,q^2,\cdots, q^N)$ as a function of $x$ and $t$. They determine the equilibrium  state of  fluid cells. In turn, the local  potentials  $\underline{\beta}\equiv (\beta^1,\beta ^2,\cdots, \beta ^N)$ are determined too once the cells reach  local time-space equilibrium.
Subsequently, one can define the average values of local conserved charges and their corresponding currents $j\equiv (j^1,j^2,\cdots, j^N)$  via
\begin{eqnarray}
q^i(x,t)& =&\langle\hat{q}^i(x,t)\rangle=\langle\hat{q}^i\rangle_{\underline{\beta} (x,t)},\\
j^i(x,t) & =&\langle\hat{j}^i(x,t)\rangle=\langle\hat{j}^i\rangle_{\underline{\beta} (x,t)},
\end{eqnarray}
becoming the Gibbs average of local densities and currents under the time and coordinates dependent potentials, respectively.

According to the GHD approach, we assume that all the local conserved quantities satisfy the continuity equations
\begin{eqnarray}
&&\partial_tq^i(x,t)+\partial_xj^i(x,t)=0
\label{continuityequations}
\end{eqnarray}
with $\qquad i=1, 2,\cdots, N$.
The density  $q^i(x,t) $ and  current $j^i(x,t)$ of the $i$-th conserved quantity 
essentially describe  variations of thermally average conservation laws at  the Euler scale.

In order to build up the description of the GHD for integrable models, it is a useful way to express the charge densities  in terms of the quasiparticle rapidity densities of the Bethe ansatz solvable models.
Here we just formulate the conserved charge densities by the quasiparticle density $\rho_p(\theta)$   for the Lieb-Liniger model, namely
\begin{eqnarray}
&&q^i(x,t)=\int d\theta \, h_i(\theta)\rho_p(\theta,x,t), \nonumber
\label{densities}
\end{eqnarray}
where $\theta$ is the corresponding rapidity,  and  $h_i$ is one particle eigenvalue of $i$-th conserved charge.
For example, particle density $q^0=\int d\theta \rho_p(\theta) $, energy density $q^1=\int d\theta \rho_p(\theta) E (\theta)  $ and momentum density $q^2=\int d\theta \rho_p(\theta)p(\theta)  $, etc.
It is also very useful to introduce the state density $\rho_s(\theta)$ and the quasiparticle occupation number $n(\theta)=\rho_p(\theta)/\rho_s(\theta)$
Here the state density $\rho_s(\theta)=\rho_p(\theta) +\rho_h(\theta) $ satisfies the Lieb-Liniger equation
\begin{equation}
2\pi \rho_s(\theta)=p{'}(\theta) +\int d\alpha \phi(\theta -\alpha ) \rho_p(\alpha),\label{BA-Lieb-Liniger-rho}
\end{equation}
which was given by Eq.  (\ref{BA-PH}). Here $p{'}(\theta)$ is the derivative of the momentum $p(\theta)$  with respect to $\theta$ and $\phi(\theta)= 2c/(\theta^2 +c^2)$ is the integral kernel function.
By properly choosing a linear space of pseudo-conserved charges as a function space spanned by all $h_i$s, one can prove $\beta^i$s form the generalized inverse temperatures or chemical potentials which are determined by the initial states.  Following the unified notations used in
 \cite{Doyon:2017-L,Doyon:2020-L}, we may express  the conserve charge $Q[h]=\sum_i\beta^i h_i (\theta) $ as a linear function of the one particle eigenvalue  $h_i(\theta)$. We define the one-particle eigenvalue $w(\theta) =\sum_i\beta_i h_i (\theta )$ of $Q[h]$, thus the quasiparticle occupation number  can  be given by the GGE states 
 \begin{eqnarray}
 n(\theta)& =&\frac{1}{1+e^{\epsilon_w} },\\
 \epsilon_w(\theta)&=&w(\theta)-\int \frac{d\gamma }{2\pi}\phi (\theta -\gamma)\log \left(1+e^{-\epsilon_w(\gamma)} \right).
 \end{eqnarray}

 Based on the analysis  given in  \cite{Castro:2016},  the energy, momentum and currents  can be formally denoted as functions of $h(\theta) $,
 \begin{eqnarray}
 \langle q[h] \rangle =\int d\theta \rho_p(\theta) h(\theta), \quad  \langle j[h] \rangle =\int d\theta \rho_c(\theta) h(\theta).
 \end{eqnarray}
 Here $ \rho_p(\theta)$  and $ \rho_c(\theta)$  are the quasi-particle density and current spectral density, respectively.
 We note that for the Lieb-Liniger model,  there is no internal degree of freedom.
 Thus the current spectral density  can be given by
 \begin{equation}
\rho_c (\theta) =:v^{\rm eff}(\theta) \rho_p(\theta),
\end{equation}
  see  \cite{Castro:2016,Bertini:2016}.
It turns out that
 \begin{eqnarray}
v^{\rm eff}(\theta) & := & \frac{\rho_c (\theta)}{ \rho_p(\theta)}=\frac{E'(\theta )+\int d\alpha \phi (\theta-\alpha ) \rho_c (\alpha ) }{p'(\theta )+\int d\alpha \phi (\theta-\alpha ) \rho_c (\alpha )}, \label{v-eff}
\end{eqnarray}
where $E(\theta)$ and $p(\theta)$ are the energy and momentum of quasiparticles in the rapidity $\theta$ space.
This equation can be rewritten as
\begin{eqnarray}
v^{\rm eff}(\theta) =v^{\rm gr}(\theta) +\int d\alpha \frac{\phi (\theta-\alpha) \rho_c(\alpha ) }{p'(\theta )}\left( v^{\rm eff}(\alpha) -v^{\rm eff }(\theta)  \right)
\end{eqnarray}
where the group velocity $v^{\rm gr}(\theta)=E'(\theta)/p'(\theta)$. In these equations of states, all conserved quantities can be obtained by the quasiparticle densities $\rho_p(\theta)$, 
i.e.
\begin{eqnarray}
q^i &=&\int \frac{d p(\theta) }{2\pi } n(\theta) h_i^{\rm dr }\nonumber\\
&& = \int d\theta \rho_p(\theta ) h_i (\theta) =p'\cdot \hat{U} h_i(\theta),\nonumber\\
j^i &=&\int \frac{d E(\theta) }{2\pi } n(\theta) h_i^{\rm dr }\nonumber\\
&&=\frac{1}{2\pi }\int d\theta n(\theta )(1-\psi \hat{\cal N} )^{-1} p'  (\theta) h^i(\theta) =E'\cdot \hat{U} h_i(\theta), \nonumber
\end{eqnarray}
here  the dressing operation is defined 
\begin{equation}
h_i^{\rm dr }=h_i(\theta)+\int \frac{d\gamma }{2\pi}\phi(\theta-\gamma)n(\gamma) h^{\rm dr}_i(\gamma).
\end{equation} 
In the above equations,
the following notations were used
\begin{eqnarray}
{\cal N} &=&2\pi n(\theta) \delta (\theta-\alpha),\quad U={\cal N}(1-\psi {\cal N} )^{-1},\nonumber\\
a\cdot b&=&\frac{1}{2\pi } \int d\theta a(\theta) b(\theta) \nonumber.
\end{eqnarray}
Following the argument on the completeness of the set of functions $h_i(\theta)$s   \cite{Castro:2016,Doyon:2017-L,Doyon:2020-L},
we have 
\begin{equation}
\int d\theta \left[ \partial _t \rho_p(\theta) +\partial _x \rho_c (\theta ) \right]h_i (\theta )=0.
\end{equation} 
Thus the GHD equations can be expressed as in different forms, for example
\begin{eqnarray}
&&\partial_t \rho_p (\theta )+\partial_x \left(v^{\rm eff} (\theta ) \rho_p(\theta ) \right)=0,\nonumber\\
&&\partial_t n(\theta )+v^{\rm eff} (\theta ) \partial_x n(\theta ) =0.\nonumber\\
&&\partial_ts(\theta ) +\partial \left( v^{\rm eff} (\theta) s(\theta) \right) =0.
\end{eqnarray}
 Here the von Neumann entropy of local fluid cell GGE states  is defined by
 \begin{equation}
 s(\theta):=\rho_s(\theta) \ln \rho_s(\theta )-\rho_p(\theta) \ln \rho_p(\theta )-\rho_h(\theta) \ln \rho_h(\theta ).\nonumber
 \end{equation}

Recently, more subtle GHD for the systems with inhomogeneous trapping potentials, quantum fluctuations and diffusion were studied theoretically \cite{Berg:2016,Caux:2019,Nardis:2018,Bastianello:2019,Bastianello:2020,Ruggierro:2020,Doyon:2017,Fava:2021,Bastianello:2022,Bouchoule:2022,Moller:2022,Bertini:2022,Scopa:2021} and experimentally \cite{Schemmer:2019,Malvania:2021,Moller:2021,Li-Chen:2020,Cataldini:2021}.

\section{IV. Experiments with quantum integrability }

In quasi-1D ultracold atomic  experiment, the particles  are tightly confined in two transverse directions and weakly confined in the axial direction.
 The transverse excitations are fully suppressed by the tight confinements.
 Thus the atoms in these waveguides can be effectively characterised by a quasi-1D system  \cite{Olshanii_PRL_1998,Dunjko_PRL_2001,Olshanii_PRL_2003}.
 In such a way, these 1D many-body systems ultimately relate to the integrable models of interacting bosons and  fermions.
We arguably say that the Lieb-Liniger gas is the most studied exactly solvable model in terms of ultracold atoms, see review \cite{Cazalilla:2011,Guan:2013,Mistakidis:2022}.
Particularly striking features involved the measurements of Lieb-Liniger gas \cite{Lieb-Liniger:1963}
 were seen in a variety of physics from ground state properties
\cite{Kinoshita:2004,Paredes:2004,Kinoshita:2005} to the thermodynamics \cite{Amerongen:2008,Armijo:2010,Jacqmin:2011,Stimming:2010,Armijo:2012,Vogler:2013,Haller:2009,Kao:2021},  quantum dynamics \cite{Langen:2015,Schemmer:2019,Malvania:2021,Kinoshita:2006,Hofferberth:2007,Ronzheimer:2013,Schweigler:2017,Prufer:2017},  the quantum GHD \cite{Schemmer:2019,Malvania:2021}, quantum criticality and Luttinger liquid \cite{Haller:2010,Meinert:2015,Yang:2017}.
On the other hand,  the Yang-Gaudin model \cite{Yang:1967,Gaudin:1967} and the  1D SU(N) Fermi gases \cite{Sutherland:1968} have also become a paradigm of ultracold fermionic atoms in 1D \cite{Liao:2010,Zurn:2012,Zurn:2013,Zurn:2013b,Wenz:2013,Murmann:2015,Revelle:2016,Pagano:2014,Song:2020}. In this section we would like to review briefly several recent experiments on quantum integrable models.

\subsection
{IV.1. Quantum Newton's Cradle}

\begin{figure}[th]
 \begin{center}
 \includegraphics[width=0.79\linewidth]{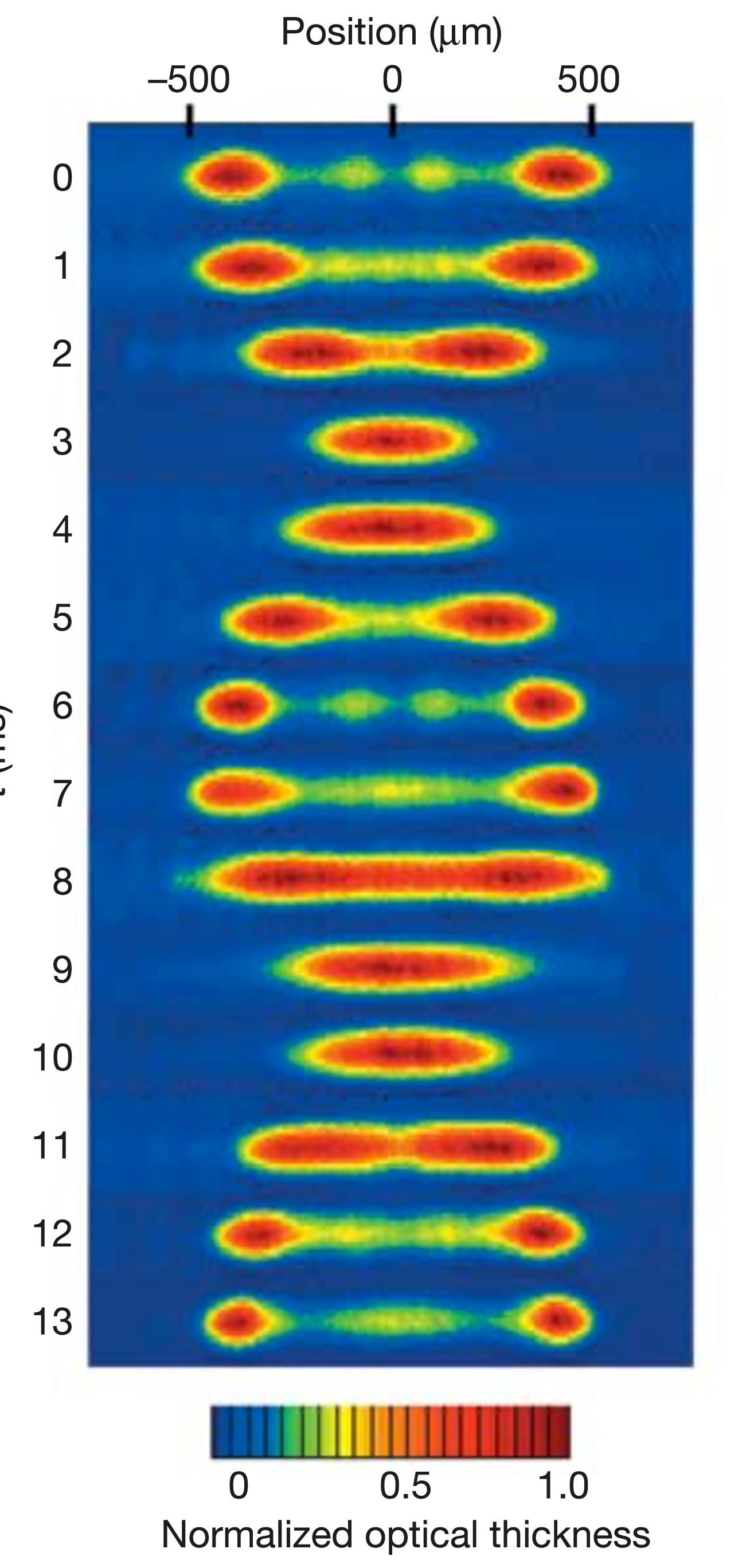}
 \end{center}
 \caption{ Absorption images read off the momentum distributions of the quasi-1D trapped Bose gas in the first oscillation cycle. The trapped gas was imparted by the  grating pulses in a superposition of $\pm \hbar k$ momentum. The two separated clouds  moved forth and back without noticeable equilibration.     Figure from \cite{Kinoshita:2006}. }
 \label{fig:Quantum-cradle-momentum}
\end{figure}

The understanding of the quantum Newton's cradle observed in \cite{Kinoshita:2006} has been received significant theoretical attention \cite{Berg:2016,Caux:2019,Nardis:2018,Bastianello:2019,Bastianello:2020,Ruggierro:2020,Doyon:2017,Fava:2021}.
In these works a key point was made that one can describe the motion of quasiparticles in the quasi-1D trapped gases after a sudden change of the longitudinal potential.
In the experiment \cite{Kinoshita:2006}, a 2D array of 1D tubes were made by a tight transverse confinement and a crossed dipole trap providing weak axial trapping. 
They measured  the density profile $f(p_{\rm ex})$  after a release of the 1D trapping potential. The 1D spatial distribution corresponds to the momentum distribution after a time-of-flight (TOF). 
They observed a separation of two clouds in the ensemble of quasi-1D trapped Bose gases of ${}^{87}$Rb atoms  over the thousands of collisions between the clouds.  In their experimental setting, each tube contained $40$ to $250$ atoms in the 2D arrays.
They integrated the images inverse to the tubes to get 1D special distributions of momentum $f(p_{\rm ex})$, see Fig.~\ref{fig:Quantum-cradle-momentum}.
From the distribution $f(p_{\rm ex})$ in space at different times, oscillations of two separated clouds showed a nearly Newton's cradle dynamics from time to time in each measurement.

 \begin{widetext}

\begin{figure}[th]
 \begin{center}
 \includegraphics[width=0.9\linewidth]{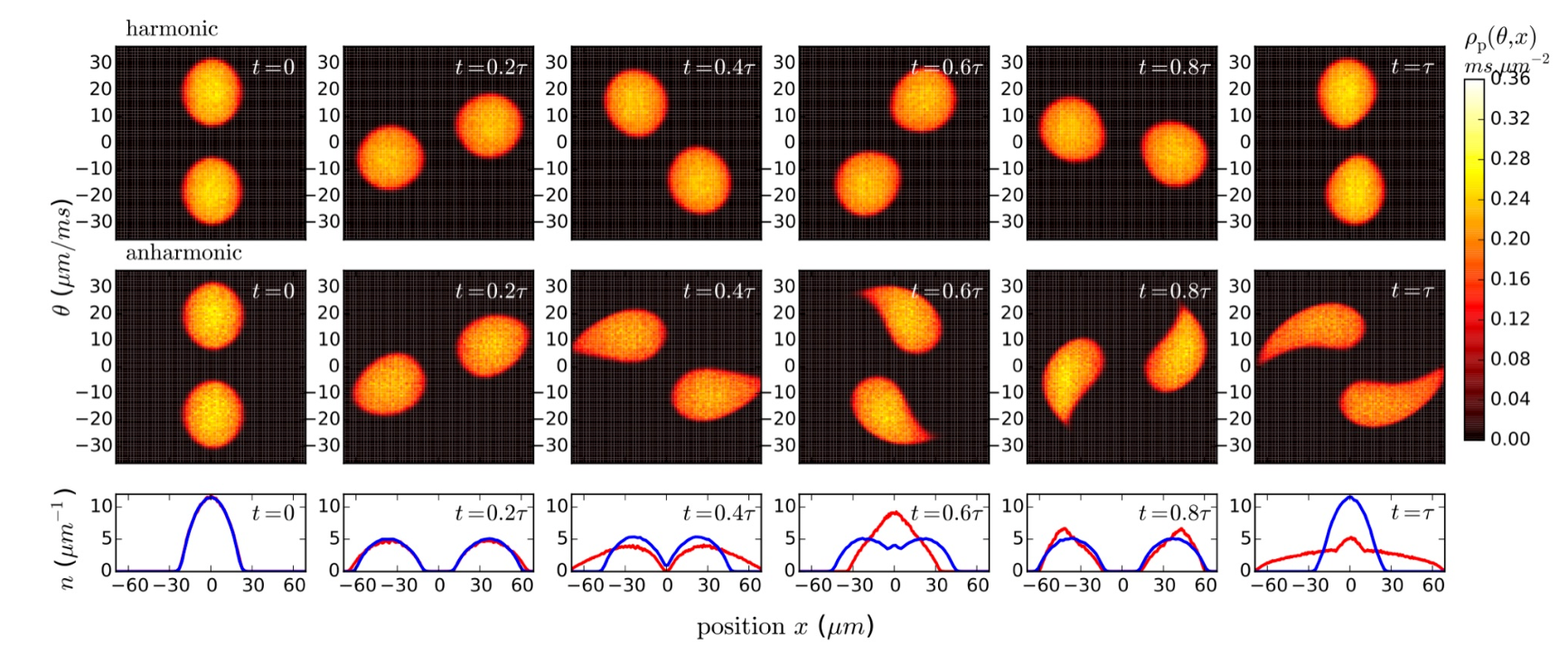}
 \end{center}
 \caption{ Demonstration of the quasiparticle density $\rho_p(\theta,x,t)$ for the quasi-1D trapped Lieb-Liniger gas in the Newton's cradle setup. The top and second rows show the time evolution of the quasiparticle density in $\theta-x$ plane within an harmonic and weakly anharmonic traps, respectively.  The bottom row  shows the  occupation number $ n(x,t) $ in $x$-{axis} at different times,  manifesting the quantum Newton-likened dynamics for the harmonic and weakly anharmonic traps.  Figure from \cite{Caux:2019}. }
 \label{fig:Quantum-cradle-GHD-S1}
\end{figure}

\end{widetext}

The quantum Newton's cradle observed in \cite{Kinoshita:2006} received  an immediate theoretical attention \cite{Rigol:2007,Ilievski:2015}. In a theoretical paper \cite{Caux:2019}, the authors gave  a quantitative   demonstration of the emergence of the quantum dynamics of Newton's cradle through numerical study of the time evolution of the quasi-1D trapped Lieb-Liniger gas with an initial state
\begin{equation}
\rho_p(\theta, x,t=0)=\frac{1}{2}\rho_p(\theta+\theta_{\rm Bragg}, x)+\frac{1}{2}\rho_p(\theta-\theta_{\rm Bragg}, x).\label{LL-GHD-initial}
\end{equation}
Where the distribution function of quasiparticles $\rho_p(\theta, x,t)$ is space-time dependent. 
The momentum of quasiparticle was  kicked by $\pm m\theta_{\rm Bragg}$ with equal probability.
The gas was  imparted into two clouds which had  a strong inter-cloud repulsion.
Based on such quantum Newton's cradle setup, the GHD equation was  given by
\begin{equation}
\partial_t \rho_p(\theta,x,t) +\partial _x \left[ v^{\rm eff}\rho_p(\theta,x,t)  \right]=\left(\frac{\partial xV(x) }{m} \right)\partial_\theta \rho_p(\theta,x,t) \label{GHD-trap}
\end{equation}
that remarkably  gave  the similar Newton's cradle dynamics as observed in \cite{Kinoshita:2006}.
In the above equation (\ref{GHD-trap}), $V(x)$ denotes the trapping potential.
In Fig.~\ref{fig:Quantum-cradle-GHD-S1},  they showed evolution of the density of quasiparticles $\rho_p(\theta,x,t)$ determined by Eq.(\ref{GHD-trap}) with the initial state (\ref{LL-GHD-initial}) using the experimental setting given in \cite{Kinoshita:2006}.
 In Fig.~\ref{fig:Quantum-cradle-GHD-S1} (bottom panel), the boson density profiles  show a dramatical oscillation of the two atomic clouds which were initially imparted by different kicks.
The real-space density profile $n(x)=\frac{1}{t_{\rm TOF}}\int d{x'} \rho_p(\theta,x^{'})$ with $\theta =x/t_{\rm TOF}$ was  directly red off from the quasimomentum distribution function of  quasi-particles $\rho_p(\theta,x)$.
The two blobs, which are separated in momentum space, evolve around the origin of phase space.
Thus the GHD provided  a full evolving  phase-space ($\theta-x$) distributions of quasiparticles associated with the Lieb-Liniger model, see upper panels in Fig.~\ref{fig:Quantum-cradle-GHD-S1}.
Here it was showed that  the GHD approach describes well slow variations of densities of particles, energy and higher conserved quantities on Euler scale.  A more detailed theoretical analysis was given in \cite{Caux:2019}.

\subsection{ IV.2. Generalized Hydrodynamics  in Lieb-Liniger Model}

\begin{figure}[th]
 \begin{center}
 \includegraphics[width=0.9\linewidth]{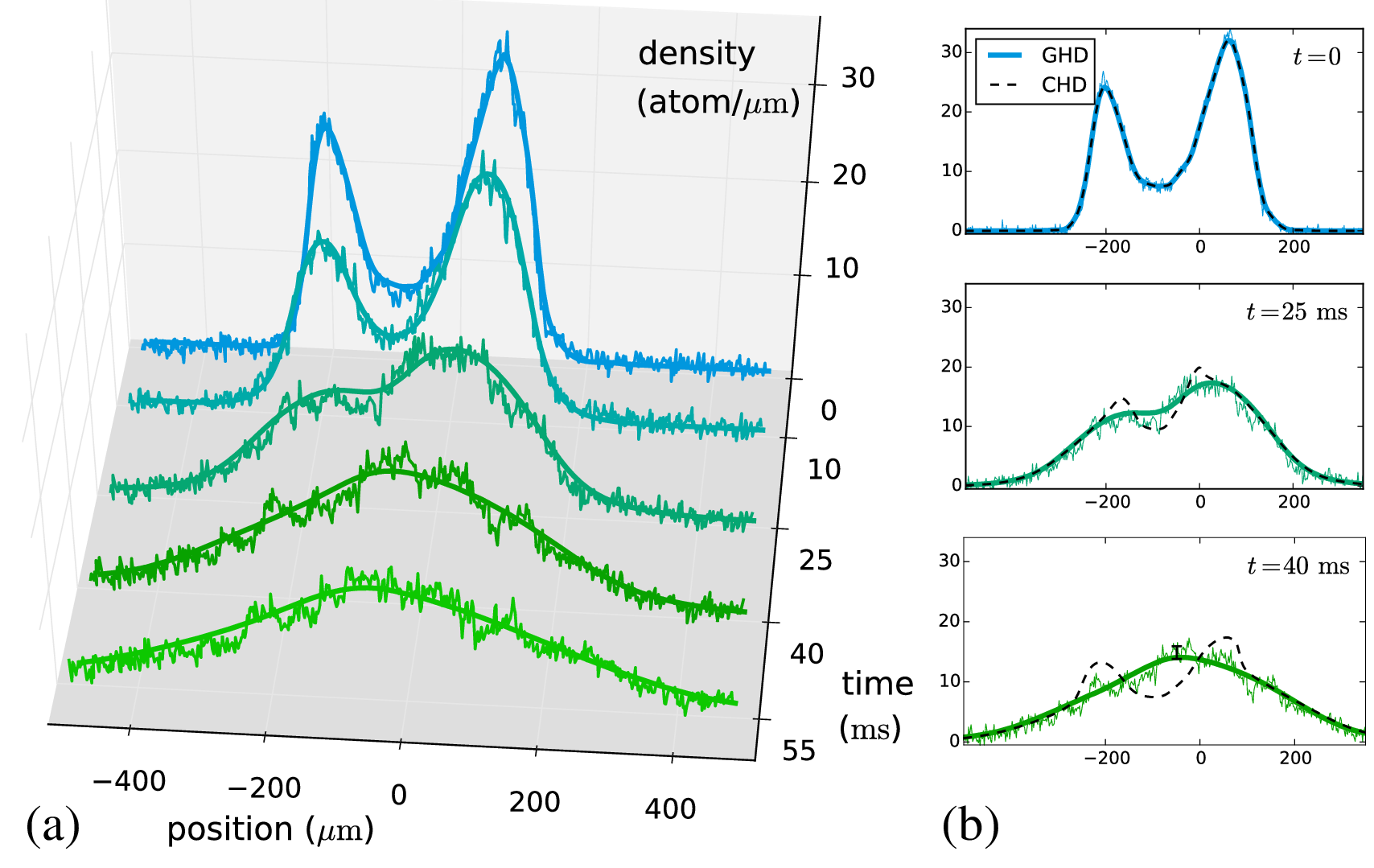}
 \end{center}
 \caption{  (a) Longitudinal expansion dynamics of a  cloud of 1D Bose gas initially trapped in a double well separated density peaks was seen in agreement with the prediction from the GHD (\ref{GHD-trap}). (b) The Longitudinal expansion profiles of a  cloud of 1D Bose gas initially trapped in a double well were compared with both the results from the GHD and CHD. This shows a clear difference between the GHD and CHD descriptions for the system.   Figure from \cite{Schemmer:2019}. }
 \label{fig:Quantum-GHD-S2}
\end{figure}

 In 2019, Schemmer {\em et al.} presented a beautiful study on the emergence of the GHD of integrable Lieb-Liniger gas on an atom chip \cite{Schemmer:2019}.
They observed time evolution of the {\em in situ} density profiles of a single 1D cloud of ${}^{87}$Rb atoms trapped on an atom chip for several different changes of longitudinal initial potentials.
The experimental data were compared with the GHD  of  trapped gas Eq. (\ref{GHD-trap}) and the conventional hydrodynamics (CHD) determined by the following equations
\begin{eqnarray}
&&\partial_t n+ \partial_x (u n)=0,\nonumber\\
&&\partial_t (m n u)+ \partial _x\left( mnu^2 +P \right)=-n\partial_x V,\nonumber\\
&&\partial_t E +\partial _x \left( u E + u P\right) =0, \label{CHD}
\end{eqnarray}
here the energy is denoted by $E=n m u^2/2+n e +nV$, the energy per particle was  denoted by $e(x,t)$,  $m$ denoted  the mass of  particles, and $u, \, n, \,P$  denoted  the velocity, particle density and pressure, respectively. They are  the functions of time $t$ and space $x$. The pressure $P$ can be  obtained from the thermodynamics Bethe ansatz equations (\ref{TBA-Bose}). Again, $V(x) $ denoted  the trapping potential.
Building on the GHD description of quantum quench dynamics in the quasi-1D trapped gases, they found a good agreement between the experimental results and theoretical simulations.
In contrast, the CHD description either works or not for 1D integrable models depending solely on smoothness of the initial trapping potential.
From the {\em in situ} density profiles they found that both GHD and CHD describe well longitudinal expansion dynamics from a harmonic trap of a 1D cloud of ultracold ${}^{87}$Rb atoms.
However, the expansion dynamics from a double well  potential further confirmed the validity of GHD for the integrable model.
It appeared to be an
 obvious discrepancy between the experimental {\em in situ} density profiles and theoretical simulation from the CHD, see Fig.~
\ref{fig:Quantum-GHD-S2}.
They demonstrated that the GHD is applicable to all interacting regimes in the quench dynamics of integrable Lieb-Liniger Bose gas on the Euler scale.

\begin{figure}[th]
 \begin{center}
 \includegraphics[width=0.9\linewidth]{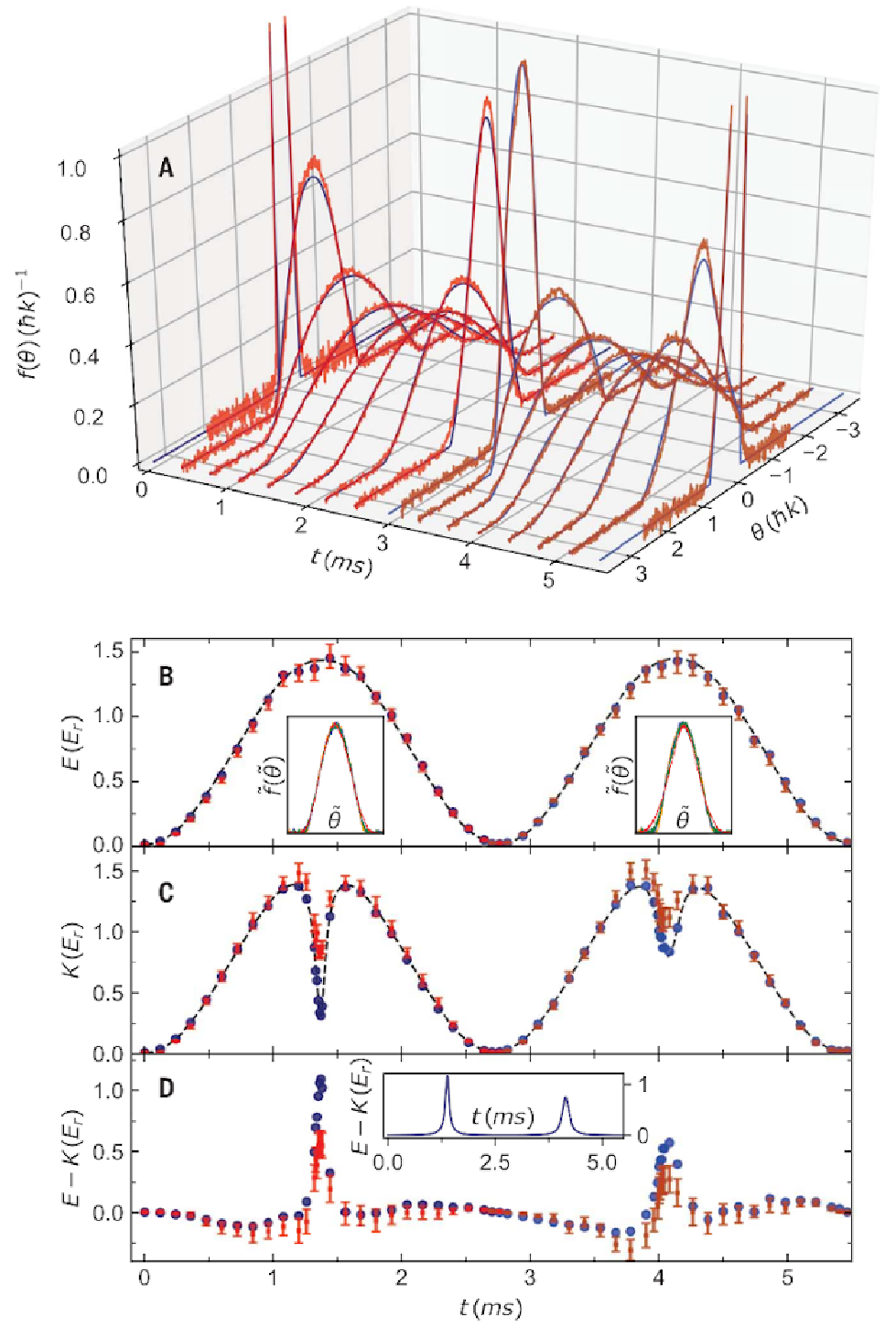}
 \end{center}
 \caption{  (A) The time evolution of the rapidity distribution after a sudden quench interaction in the first two cycles. The experimental distribution curves agree well with the blue ones predicted from the GHD theory by counting the particle numbers in each tubes, see Eq.(\ref{f-theta}).
 (B) The time evolution of the weighted rapidity energy $E$  after the  sudden quench interaction in the first two cycles show a good agreement between experiment data (red squares) and GHD theory (blue circles), see Eq.(\ref{E-t}). The two insets show the rescaled rapidity distribution for four phases  points near $0,\,\pi/4,\, \pi/2,\,3\pi/4,\,\pi$ (in different color) in the two cycles. The distributions are no longer self-similar in the second cycle.
 (C) and (D) respectively show the time evolutions of the weighted kinetic energy $K$ and the interaction interaction $E_{\rm I}=E-K$
 after the sudden quench interaction in the first two cycles.  They show a good agreement between experiment data (red squares) and GHD theory (blue circles).  see Eq.(\ref{K-t}) and Eq.(\ref{EI-t}). Figure from \cite{Malvania:2021}. }
 \label{fig:Quantum-GHD-S3}
\end{figure}

In 2021, D. S. Weiss's group \cite{Malvania:2021} further demonstrated the GHD of the quasi-1D trapped Bose gas in both the strong and intermediate coupling regimes.
They measured the rapidity distributions $f(\theta) $ in several ways by changing the quench potentials.
Such changes  of quench were still small enough so that the transverse excitations were fully suppressed.
The key result of their paper was the validation of the GHD via the (per particle) rapidity momentum distribution $f(\theta)$, the ( per particle) rapidity energy $E(t) $, the (per particle) kinetic  energy  $K(t)$ and (per particle)  interaction energy $E_{\rm I}(t) $, which are given by the Bethe ansatz rapidity
\begin{eqnarray}
f(\theta, t) &=& \frac{1}{N_{\rm tot}(t)  } \sum_\ell \int d z \rho_\ell (\theta, z, t),\label{f-theta} \\
E(\theta, t) &=&  \frac{1}{N_{\rm tot}(t)  } \sum_\ell \int d z \rho_\ell (\theta, z, t)\left[\frac{\theta }{m} -v^{\rm eff}_\ell \right] \theta,\label{E-t}\\
K(\theta, t) &=&  \frac{1}{N_{\rm tot}(t)  } \sum_\ell \int d z \rho_\ell (\theta, z, t)\left[v^{\rm eff}_\ell -\frac{\theta }{2m}\right] \theta,\label{K-t}\\
E_{\rm I} (\theta, t)&=&  \frac{1}{N_{\rm tot}(t)  } \sum_\ell \int d z \rho_\ell (\theta, z, t)\frac{\theta^2 }{2m},\label{EI-t}
\end{eqnarray}
where $ \rho_\ell (\theta, z, t)$ is the local density of quasi-particles in the $\ell$-th tube, the sum runs over all tubes in the 2D arrays.
In each fluid cell the rapidity density $\rho(\theta ) =\frac{1}{L}\sum _{i=1}^{N_{\rm cell }}\delta (\theta -\theta_j)$ satisfies the local rapidity Bethe ansatz equation (\ref{BA-Lieb-Liniger-rho}). The above formula can be derived in a straightforward way via the Bethe ansatz equation.
Considering zero entropy limit, the measured evolutions of the above quantities in a sudden quench of  the interaction from an initial state were seen in agreement with the prediction of the GHD theory, see Fig.~\ref{fig:Quantum-GHD-S3}.

From the above discussion, we see that the isolated integrable systems in  general  do not thermalize during evolution from an initial quench. The thermalization near integrability in the dipolar quantum Newton's cradle was recently studied \cite{Tang:2018}.
In this system, the 1D Bose gas of Dysprosium atoms   with a strong magnetic dipole-dipole interaction was created by tuning the strength of the integrability-breaking perturbation and nearly integrability. They found a strong evidence for thermalization close to a strongly interacting integrable point occurred followed by near exponential thermalization. The measured thermalization rate is consistent with theoretical simulation. Integrable and non-integrable quantum systems of ultracold atoms with tunable interactions open a new venue to study thermalization and thermodynamics in and out-of-equilibrium.

\subsection{IV.3.  Dynamical Fermionization}

On the other hand, the Bose-Fermi mapping mechanism   \cite{Girardeau:1960,Girardeau:1965} suggests that the physical properties like density profile, the thermodynamical properties and the density and density correlation are the same for both the ideal Fermi gas and the  homogeneous 1D Tonks-Girardeau  Bose gas.
However, the physical quantities related to the one-body density matrix with off-diagonal elements, consequently the momentum distribution for the Tonks-Girardeau  Bose gas significantly  differ from the one for the ideal Fermi gas.  This is mainly because of the quantum statistics play a significant role in correlations. The dynamics of the  Tonks-Girardeau  Bose gas \cite{Minguzzi:2005,Rigol:2005} led  to the notion of dynamical fermionization.
The dynamical fermionization of the 1D Tonks-Girardeau Bose gas
 reveals the typical expansion dynamics produced by the many-body wave function \cite{Lyer:2012,Buljan:2008,Campbell:2015}.

The strongly interacting bosonic Tonks-Girardeau gas was initially trapped in harmonic potential. After a sudden release of the axial confinement, the momentum distribution of the gas rapidly evolves to that of the ideal Fermi gas in the initial trap. This phenomenon \cite{Minguzzi:2005,Rigol:2005} is now referred to the dynamical fermionization.
It has been theoretically demonstrated for several integrable models of ultracold atoms, such as 1D Bose gas \cite{Bolech:2012,Campbell:2015,Xu:2017}, 1D anyon gas \cite{del-Campo:2008}, 1D spinor Bose gas \cite{Alam:2021}, and 1D Bose-Fermi mixture and anyons  \cite{Patu:2021,Piroli:2017}.
Now it is well established that the wave function of the Lieb-Liniger gas (\ref{Ham1}) with an infinity strong repulsion can be written in terms of the one of the ideal fermions with the antisymmetric function in the same potential, referred as Bose-Fermi mapping \cite{Girardeau:1965,Girardeau:1960}.
In a 1D harmonic trap $V(x,t)=m\omega^2(t)x^2/2$ with a time dependent trapping frequency $\omega(t)$ and let $\omega(t\le 0)=\omega_0$, the many-body wave function of the the Tonk-Girardeau gas is given by
\begin{equation}
\Phi (x_1,\ldots x_N;t)=A(x_1,\dots,x_N)\Phi_F (x_1,\ldots, x_N;t), \label{wave-function-TG}
\end{equation}
where $A(x_1,\dots,x_N)=\Pi_{1\le j<k\le N}sgn (x_j-x_k) $ is an antisymmetric function and  the non-interacting fermion wave function is given by
\begin{equation}
\Phi_F (x_1,\ldots, x_N;t)=\frac{1}{\sqrt{N!} }{\rm det } _{j,k=1}^N\phi (x_k,t).
\end{equation}
The  single-particle wave function $\phi (x_k,t)$ can be obtained by solving the time-dependent Schr\"{o}dinger equation with the following solution
\begin{equation}
\phi (x_k,t)=\frac{1}{\sqrt{b}}\phi_j\left( \frac{x}{b},0\right)\exp\left(\mathrm{i} \frac{mx^2}{2\hbar }\frac{\dot b}{b}  -\mathrm{i} E_j\tau (t)\right),
\end{equation}
where $E_j$ is the energy of the $j$th single-particle eigenstate of the initial  harmonic trap, the scaling parameter  $b(t)$ obeys the second differential equation $\ddot b+\omega ^2(t) b =\omega _0^2/b^3$ with the initial conditions $b(0)=1$ and $\dot b (0)=0$. While the time scaling
parameter $\tau (t)=\int^t_0d t'b^{-2}(t')$, and $\phi_j(x,0)$ is the wave function of the 1D harmonic oscillator with frequency $\omega_0$ and energy $E_j$,
\begin{equation}
\phi_j(x,0)=\frac{1}{\sqrt{ 2^j j! \sqrt{\pi} }}\left( \frac{m\omega }{\hbar }\right)^{1/4}e^{-\frac{m\omega_0 x^2}{2\hbar }} H_j\left(\sqrt{\frac{m\omega_0}{\hbar } }  x\right). \nonumber
\end{equation}

\begin{figure}[th]
 \begin{center}
 \includegraphics[width=0.9\linewidth]{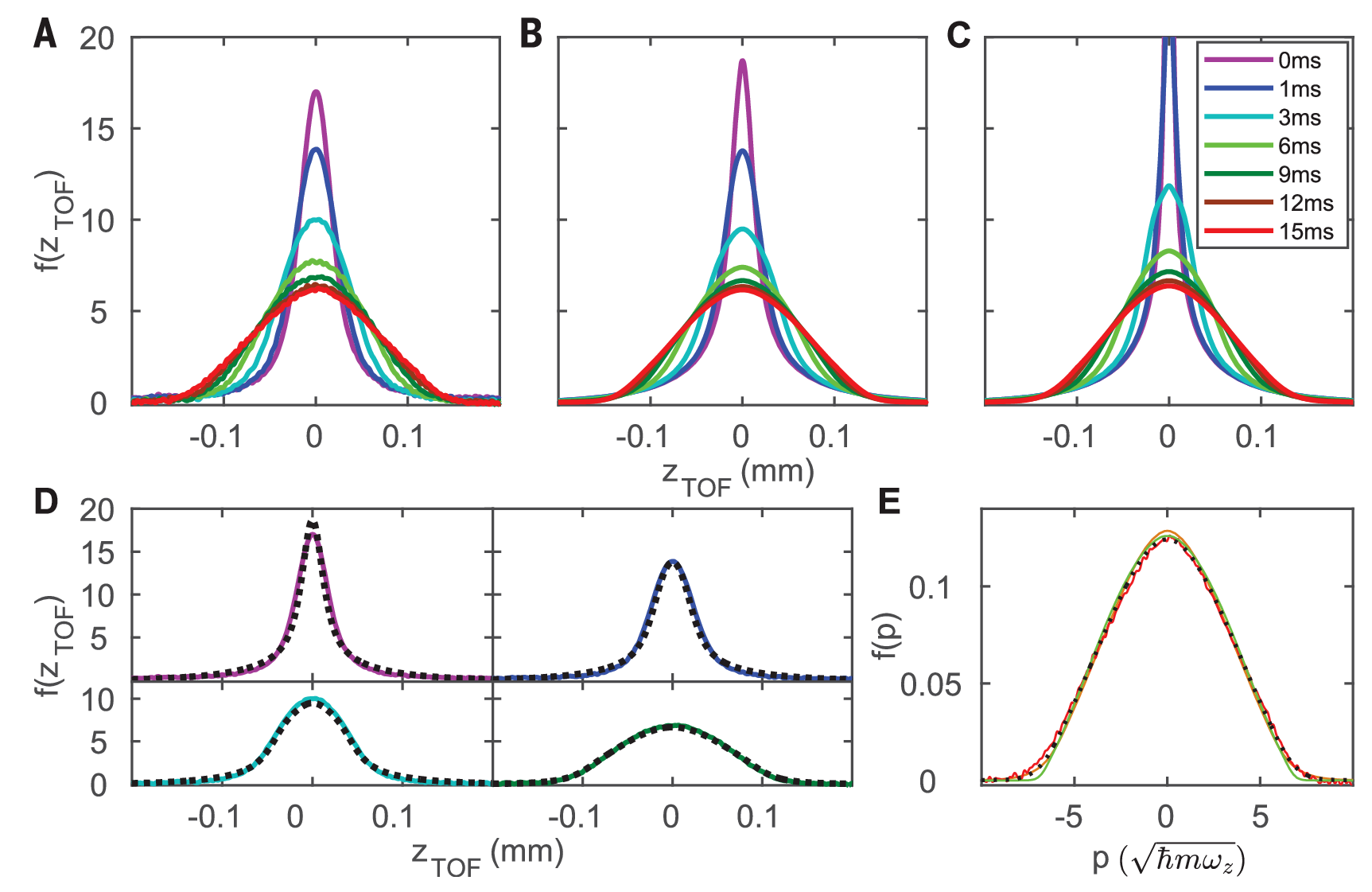}
 \end{center}
 \caption{  Dynamical fermionization of the Tonks-Girardeau gas of ${}^{87}$Rb atoms. (A) The normalized TOF axial  profiles at different expansion time. The initially peaked bosonic TOF distribution smoothly  evolves  to the rounded fermionic one over the first $12$ms. Over the $12$ms, it deforms  to the one of the ideal Fermi gas. (B) The corresponding theoretical simulation of the expansion profiles for the Tonks-Girardeau gas. (C) The theoretical simulation of the corresponding momentum distributions. (D) Experimental  TOF distributions from (A) are  compared with the corresponding theoretical simulation one from (B) at the evolution time $t_{\rm ev} =0,1,\, 3,\, 9$ms. (E) Comparison of theoretical and experimental TOF expansion  curves at $t_{\rm ev} =15$ms: Experimental (red) and theoretical (dotted black) TOF profiles agree with the momentum distributions  from the theoretical Tonks-Girardeau gas (orange) and the non-interacting Fermi gas (green).
Figure from \cite{Wilson:2020}. }
 \label{fig:Lieb-Liniger-expansion}
\end{figure}

Then the many-body wave function (\ref{wave-function-TG}) can be given explicitly by
\begin{eqnarray}
\Phi (x_1,\ldots N;t)&=&\frac{1}{b^{-N/2}} \Phi_T (x_1/b,\ldots x_N;0)\nonumber\\
&&\times e^{\frac{\mathrm{i} \dot b}{b\omega_0}\sum_j\frac{x_j^2}{2\ell_0^2} }e^{-\mathrm{i} E_j \tau},
\end{eqnarray}
where $l_0=\sqrt{\hbar /(m\omega_0)} $.
Using this many-body wave function, the one-body density matrix can be calculated analytically and numerically, namely,
\begin{eqnarray}
&&g_1(x,y,t)=\\
&&N\int dx_2\cdots dx_N\Phi^*(x,x_2,\ldots, x_N;t)\Phi(y,x_2,\ldots, x_N;t).\nonumber
\end{eqnarray}
The time evolution of the momentum distribution is given by
\begin{equation}
n(p,t)=\int dx dy e^{\mathrm{i} p(x-y) /\hbar }g_1 (x,y;t).
\end{equation}
Thus the expansion or quench dynamics of the 1D strongly interacting Bose gas can be studied by switching off the axial confinement potential or suddenly changing the trapping frequency.
The quench dynamics of the 1D Bose gas with an arbitrary interaction strength was studied  by D Iyer and N Andrei in \cite{Lyer:2012}.
They showed that for any value of the interaction strength, as long as it is repulsive, the evolution of the system in  a long time asymptotically approaches
to the one of the hard core bosons at $c\to \infty$. 
Such dynamical fermionization is due to  the structure of
the two-body $S$-matrix. The momenta integration contours in the wave function of Yudson representation are determined by the interaction.  The interacting strength $c$ can be rescaled by the time, i.e. $c\to c\sqrt{t}$ in the evolved wave function. The coupling constant effectively evolves with time.   In the long time limit, the time rescaled  scattering matrix $S^{ij} \to -1$, so that the  asymptotic wave function of the bosons was represented by the one of the  free fermions \cite{Lyer:2012} by using the stationary phase approximation.
Thus the repulsively interacting bosons turn into fermions as time evolves in a long time \cite{Buljan:2008,Campbell:2015}. 
The fermionization also occurs in 1D  lattice model of bosons and interacting anyons  \cite{Piroli:2017}.

Following theoretical work \cite{Minguzzi:2005,Rigol:2005,Bolech:2012,Campbell:2015,Xu:2017}, the observation of the dynamical fermionization was carried out in \cite{Wilson:2020}. In this experiment, it was remarkable  to observe that the momentum distribution of the Tonks-Girardeau gas evolved rapidly to the one of the ideal Fermi gas after an expansion in 1D, see Fig.~\ref{fig:Lieb-Liniger-expansion}.
In this experimental setting, the interaction between the atoms was negligible upon expansion in the flat axial potential.
The time-of-flight  expansion profiles are in good agreement with the theoretical prediction for the Tonks-Girardeau gas as discussed above.
At the evolution time $t_{\rm ev} =15$ms, the Tonks-Girardeau gas rapidity distribution is the same as the momentum distribution of the ideal Fermi gas.
The control capability   to measure the momentum and rapidity distributions allow to further investigate more subtle quantum transport properties and diffusion in 1D quantum atomic gases.

\begin{figure}[th]
 \begin{center}
 \includegraphics[width=0.9\linewidth]{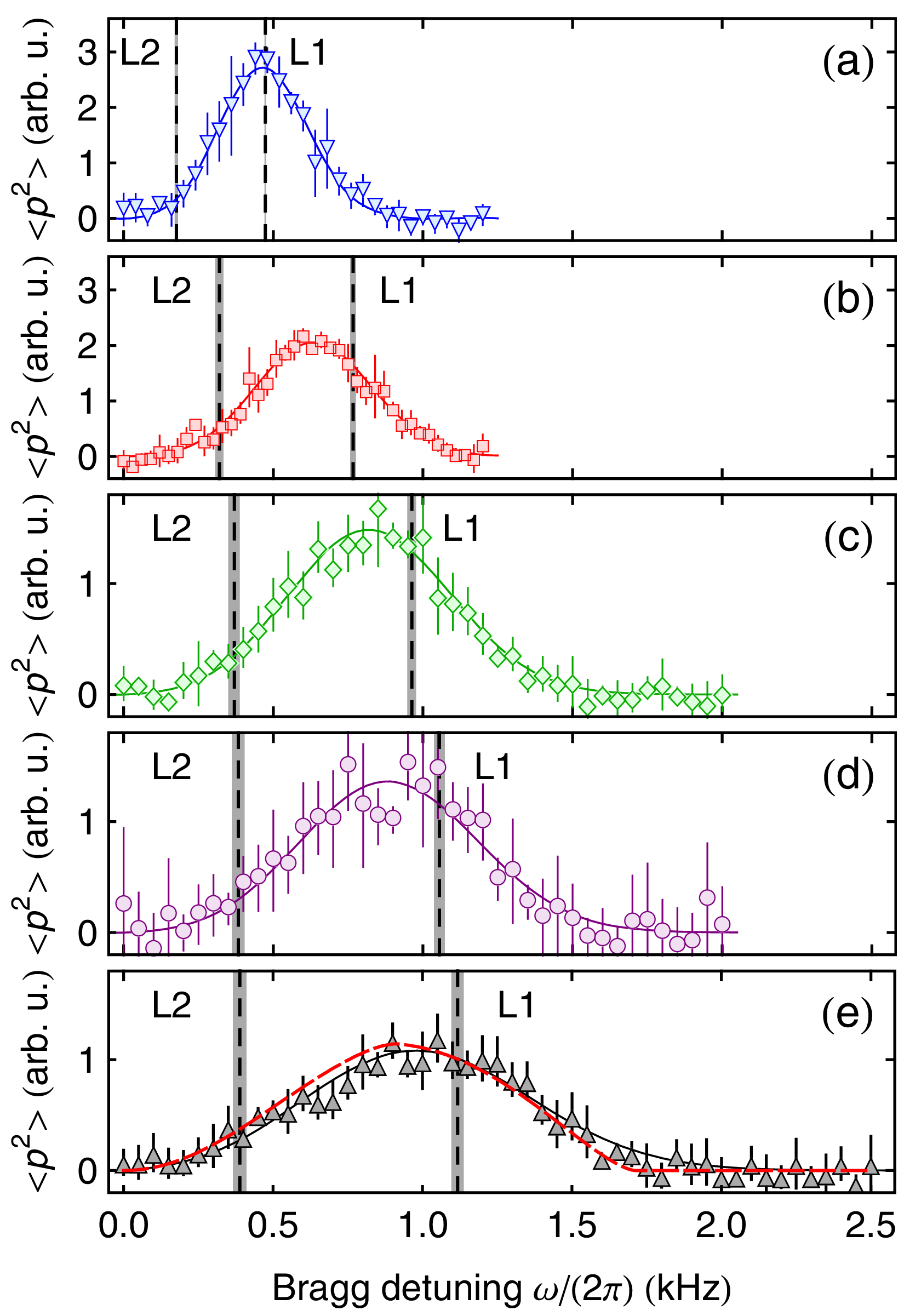}
 \end{center}
 \caption{  Bragg signal v.s. detuning frequency $\omega/ 2\pi $(kHz) for the Lieb-Liniger gas with different interaction strengths. The scattering length was set for $15a_0$ (a), $ 173a_0$ (b), $399a_0$ (c), $592a_0$ (d) and $819a_0$ (e), respectively.  The experimental Bragg signal was compared with  the Gaussian fit with a multiplication of the $\omega$ (solid lines) for (a)-(d). The vertical dashed lines stand for the positions of the Type-I particle excitation ($L_1$) and the Type-II  hole excitation ($L_2$). The shape of the experimental data in (e) fits   well the particle-hole  excitations of the trapped Tonks-Girardeau gas, showing the  signature of the typical Lieb-Liniger Type I and Type II excitations.  Figure from \cite{Meinert:2015}. } 
 \label{fig:Lieb-Liniger-excitations}
\end{figure}

\subsection{IV.4. Excitations and quantum Holonomy in Lieb-Liniger  Gases }
{\bf Elementary excitations: } The Lieb-Liniger model \cite{Lieb-Liniger:1963}  has been extensively studied in literature, see a review \cite{Cazalilla:2011}.
In the last two decades this model has provided an unpreceedent ground for experimental investigation of a wide range of many-body phenomena, see introduction part.
Regarding the low energy physics of the  Lieb-Liniger gas, experimental observations of collective excitations in the Lieb-Liniger gases have been achieved in \cite{Fabbri:2015,Meinert:2015}.
In the paper \cite{Meinert:2015}, by using the Bragg spectroscopy, the excitations for the quasi-1D trapped gases of Cesium atoms  with tunable interaction strength were studied. 
An ensemble of independent 1D tubes was created by anisotropic confinements.
The scattering length was tunable via a broad Feshbach resonance.
The measured  Bragg signal for a fixed value of momentum $k=3.24 (3) $ (about the mean value of $k_F$)  was compared with the dynamical structure factor (DSF) $S(k,\omega)=\int dx \int dt e^{\mathrm{i} \omega t-\mathrm{i} kx}\langle \rho(x,t)\rho(0,0)  \rangle $ at finite temperatures, i.e. $\delta E(k,\omega )\propto \hbar \omega \left(1-e^{-\hbar \omega /(k_B T) }\right)S(k,\omega)$.
In this setting, the Luttinger liquid prediction on the DSF for the linear dispersion was not enough.
They observed a clear broadening of the DSF peaks due to the Lieb-Liniger Type I (particle excitation $L_1$) and Type II  (hole excitation $L_2$) excitations, see Fig.~ \ref{fig:Lieb-Liniger-excitations}.
The broadening signature for both excitations was revealed from the Bragg signal.
In the figures (a)-(e), the broadening of the DSF spectral shape mainly came along with increasing  the  interaction strength.
In particular, for the Tonks-Girardeau regime in Fig.~\ref{fig:Lieb-Liniger-excitations}(e),  the experimental signals  agree  with the  theoretical simulation with  accounting for the effects of the inhomogeneous density distribution in each  tube and  the density
distribution averaged over the ensemble of tubes at a temperature. 
The DSF   was  evaluated at finite temperature via the Bethe Ansatz-based numerical method, see \cite{Meinert:2015}. 
For low temperatures, the spectra showed a band width between the Lieb-Liniger Type I and Type II excitations.
However, the band curvature effect in the DSF signal was not directly examined yet in this paper. 

The temperature played an important  role through the DSF and  the inhomogeneous density  distributions along the tubes in the theoretical simulation.
This experiment opened to study low energy excitations for the integrability-based quantum systems of ultracold atoms.

\begin{figure}[th]
 \begin{center}
 \includegraphics[width=0.99\linewidth]{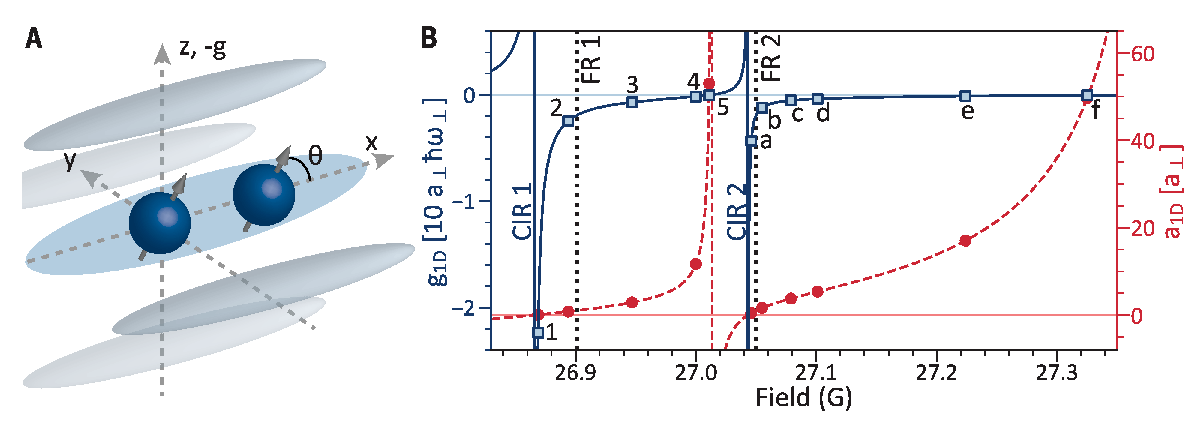}
 \end{center}
 \caption{ Experimental setup for quantum holonomy.
 (A) The schematic illustration of the arrays of 1D traps in a 2D optical lattice. The dipoles angle $\theta$ along the axial direction can be changed by external magnetic field.
 (B) The effective 1D scattering length $a_{\rm 1D}$ (red dashed lines) and interaction strength $g_{\rm 1D}$ (blue solid lines) can be tuned by the confine-induced-resonances through the applied field.  The two resonance positions (dot-dashed lines) are  showed for the two cycles of quantum holonomy.  The numbers and letters indicate the measured positions for the energy per particle shown in Fig.~\ref{fig:Lieb-Liniger-holonomy2}.
   Figure from \cite{Kao:2021}. }
\label{fig:Lieb-Liniger-holonomy1}
\end{figure}

{\bf Quantum holonomy:} Stable highly excited states of interacting systems are extremely important  in both theory and experiment.
Such a metastable   state can exist in  the strong interacting bosons in 1D, referred to the super Tonks-Girardeau  gas,
which was predicted theoretically by Astrakharchik {\em et al.} \cite{Astrakharchik:2005}  using Monte Carlo simulations and later by Batchelor, Guan and collaborators  using the integrable interacting Bose gas with attractive interactions \cite{Batchelor:2005}.
In a surprising experiment, Haller {\em et al.} \cite{Haller:2009} made a  breakthrough by observing the  stable highly excited gas-like state--the super Tonks-Girardeau gas in the strongly attractive regime of bosonic Cesium atoms.
 This observation  has improved our understanding of quantum dynamics in many-body physics \cite{Chen:2010a,Chen:2010b,Yonezawa:2013,Tang:2018}.
A physical intuition for  the super Tonks-Girardeau gas is its metastable gas-like state with a stronger Fermi-like pressure than for free fermions which prevents a collapse of atoms in the strongly attractive interaction regime \cite{Chen:2010a,Panfil:2014,Kormos:2011,Girardeau:2009}.
 From the Bethe ansatz, we observe that the energy at  the limits $c\to \pm \infty$ is analytic,
where the compressibility given by
\begin{equation}
  \frac{1}{\kappa} = 2 \pi^2 n - \frac{16 \pi^2}{c} n^2 + \frac{80 \pi^2}{c^2}
  n^3 + \left( \frac{64}{3} \pi^2 - 320 \right)  \frac{n^4}{c^3}.\label{cop-s}
\end{equation}
is alway positive.
Therefore such state is stable against collapse in the strongly attractive regime, even for the whole attractive regime.
The energy of  the super Tonks-Girardeau state can be increased  smoothly when the attraction is ramped  down to the noninteracting limit.
A next  cycle with an energy pumping can be repeated over the last cycle.
This phenomenon is now  called quantum holonomy \cite{Yonezawa:2013}.

Such  exotic quantum holonomy was surprisingly demonstrated in a recent experiment did by B L Lev's group in Stanford University \cite{Kao:2021} using the 1D Bose gas of dysprosium atoms  with dipole-dipole interactions.
The relevant Hamiltonian is  given by
\begin{eqnarray}
H&=&-\frac{\hbar^2}{2m}\sum^N_{j=1} \frac{\partial ^2}{\partial x_j^2}+\sum_{a\le i<j\le N}\left[ g_{1D} \delta(x_i-x_j)\right. \nonumber \\
&&\left.  + V^{\rm 1D}_{DDI}(\theta, x_i-x_j) \right],\label{Ham-DD}
\end{eqnarray}
where $g_{\rm 1D} (B)=2\hbar ^2/(m a_{\rm 1D}(B))$ is tunable by controlling the confine-induced-resonance. The dipole-dipole interaction $V^{1D}_{DDI}$ essentially depends on the atomic dipole aligned angle $\theta$ by applying an external magnetic field to polarize dipoles  through the scale factor $(1-3\cos^2 \theta)$, see Fig.~ \ref{fig:Lieb-Liniger-holonomy1}. In the absence of the dipole-dipole interaction, the model (\ref{Ham-DD}) reduces to  the Lieb-Liniger model (\ref{Ham1}).   In this experiment, two cycles of quantum holonomy were demonstrated through tuning the interaction strength $g_{\rm 1D}$.
 The ensemble comprises an array of about $1000$ 1D optical traps with about $30-40$ atoms per tube from the edge to the central tubes.

\begin{figure}[th]
 \begin{center}
 \includegraphics[width=0.89\linewidth]{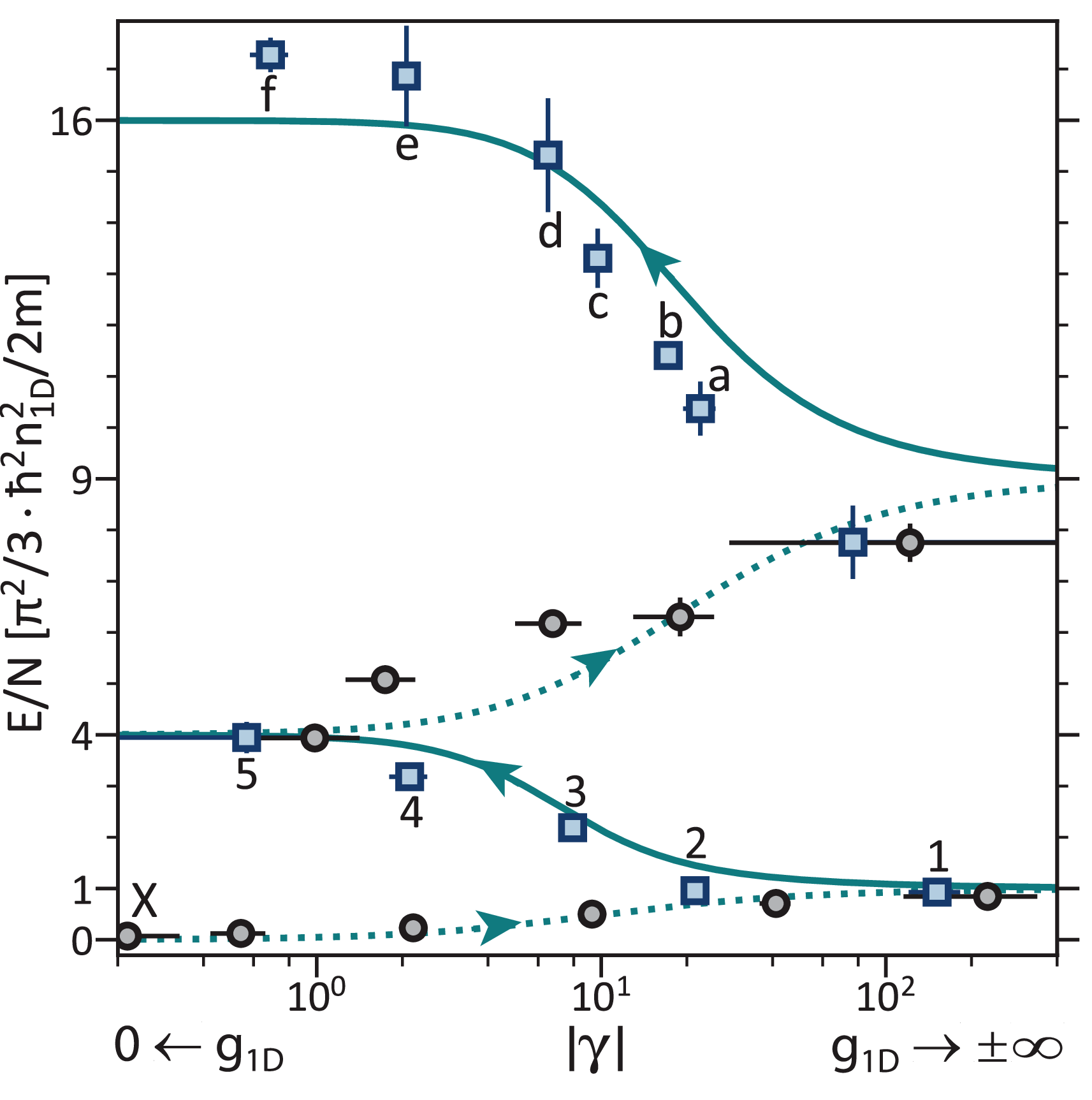}
 \end{center}
 \caption{ Energy per particle  for two cycles of quantum holonomy.
 The energies  per particle were measured for the quasi-1D trapped Bose gas of dysprosium atoms with dipole-dipole interactions at different interaction strengths in the two pumping cycles.
 The black circles (blue squares) stand for the measured positions of $g_{\rm 1D}$ in the repulsive  (attractive ) regimes in the two cycles. The dotted (solid) lines are the Bethe ansatz solution simulation  (\ref{BAE}) for the model  without the dipole-dipole interaction.  The arrows indicate the pumping direction. Different interaction strengths were indicated  by the numbers ($1,2, \cdots ,5$) and the letters ($a,b,\cdots, e$) in  Fig.~\ref{fig:Lieb-Liniger-holonomy1}.
   Figure from \cite{Kao:2021}. }
 \label{fig:Lieb-Liniger-holonomy2}
\end{figure}

This experiment elegantly created a hierarchy quantum pumping by cyclically changing the short range interacting $g_{\rm 1D}$ in the quasi-1D trapped Bose gas (\ref{Ham-DD}) for a fixed angle $\theta =90^0$, see Fig.~\ref{fig:Lieb-Liniger-holonomy2}, where the per particle energy against the dimensionless interaction strength $|\gamma|$ was measured through time-of-flight images. The authors demonstrated that the energy of the first cycle Tonks-Girardeau gas at the positive infinite repulsion limit  is identical to the one of the second cycle  super Tonks-Girardeau gas at the negative  infinite repulsion limit. In contrast to the previous experiment with Cesium atoms by  Haller {\em et al.} \cite{Haller:2009}, the dipole-dipole interaction at the "magic" angle  $\theta =90^0$ can stablize the super Tonks-Girardeau state at finitely strong attractions. Here the dipole-dipole interaction breaks the quantum integrability.  However, such quantum holonomy does not seem to restrict to the integrable model. We would like to remark that this quantum holonomy can occur in 1D Bose gases  because a full fermionization of interacting bosons at infinite strong interaction limit  happens only in 1D. So that  such highly excited states may  analytically connect to the non-interaction limit.   Nevertheless,  the super Tonks-Girardeau states can occur in 1D interacting bosons and fermions, and p-wave interacting Fermi gases \cite{Yin:2011,Imambekov:2010}.

\subsection{IV.5.  Luttinger liquids and Quantum criticality in Lieb-Liniger  Gases }

In 2011,  Guan and Batchelor \cite{Guan:2011}  calculated universal scaling functions of thermodynamics of   the Lieb-Liniger Bose gas.
The scaling behaviour of the thermodynamics of such an interacting Bose gas can be used to map out the criticality of the 1D model of  ultracold  atoms in experiment. The expression for the equation of state allows the exploration of Tomonaga-Luttinger liquid physics and quantum criticality in an arguably simple quantum system, see discussion of  Lieb-Liniger Bose gas in Section II.
At zero temperature, a quantum transition  from vacuum to the TLL occurs when  the chemical potential $\mu$ approaches to the critical point $\mu_c=0$.
 The equation of the state is given by  the universal scaling form of the density
\begin{eqnarray}
  n(T, \mu) \approx n_0 (T)+ T^{d/z + 1 - 1/\nu z} \mathcal{F}
  \bigg(\frac{\mu - \mu_{\rm c}}{T^{1/\nu z}} \bigg), \label{density-scaling}
\end{eqnarray}
where the background density $n_0 (T)\sim 0$ in the limit $T\to 0$.
The scaling function $\mathcal{F}(x) = - \frac{1}{2 \sqrt{\pi}} {\rm
Li}_{1/2} (-{\rm e}^x)$ reads off the dynamic critical exponent $z =
2$, and the correlation length exponent $\nu = 1 / 2$.
The universal scaling  functions  of compressibility $\kappa$ and specific heat $\tilde{c}_{\rm v} =c_{\rm v}/T$ are given by
\begin{eqnarray}
  \kappa &=& \kappa_0(T)  + T^{d/z + 1 - 2/{\nu z}} \mathcal{K} \bigg(
  \frac{\mu - \mu_{\rm c}}{T^{1/\nu z}} \bigg),\\
  \nonumber
   \tilde{c}_{\rm v} &=&\tilde{c_0}(T)   +T^{d/z + 1 - 2/{\nu z}} \mathcal{C} \bigg(
  \frac{\mu - \mu_{\rm c}}{T^{1/\nu z}} \bigg),
\end{eqnarray}
where the regular parts $\kappa_0 (T)=\tilde{c_0}(T) = 0$  for the system in low temperature limit and the scaling functions ${\cal K}( x) = - \frac{1}{2 \sqrt{\pi}} {\rm Li}_{-
{1}/{2}} (x)$ and ${\cal C}( x) = - \frac{3}{8 \sqrt{\pi}} {\rm Li}_{-
{1}/{2}} (x)$.

The low-lying excitations  present  the  phonon dispersion $\Delta E(p) =v_s p$ in the long-wavelength limit, i.e. $p\to 0$.
In this limit, the following   effective Hamiltonian \cite{Haldane:1981,Giamarchi-book}
\begin{eqnarray}
H=\int d x \left(\frac{\pi v_{\rm s} K}{2} \Pi^2+\frac{v_{\rm
s}}{2\pi K }\left(\partial _x \phi \right)^2\right),
\end{eqnarray}
essentially describes the low-energy physics of the Lieb--Liniger  gas,
where the canonical momenta $\Pi$ conjugate to the phase  $\phi$  obeying  the standard Bose commutation relations $\left[ \phi(x), \Pi(y) \right]=\mathrm{i} \delta(x-y)$.
$\partial_x \phi$ is proportional to the density fluctuations.
In this effective Hamiltonian, $v_s/K$ fixes the energy for the change of density.
From the Bosonization approach, the leading order of one-particle correlation $\langle{
\psi^{\dag} (x) \psi(0) } \rangle\sim 1/x^{1 / 2 K}$ is uniquely
determined by the Luttinger parameter $K$.
The Luttinger parameter is given  by the ratio of sound velocity to stiffness, namely,
\begin{eqnarray}
 K=\frac{v_{\rm s}}{v_N}
 = \frac{\pi}{\sqrt{3 e_0(\gamma) -2\gamma\frac{{\rm d}e_0(\gamma)}{{\rm d} \gamma} + \frac{1}{2} \gamma^2 \frac{{\rm d}^2  e_0(\gamma)}{{\rm d}
  \gamma^2}}},
\end{eqnarray}
where $v_{\rm s}$ is the sound velocity and $v_N$ is the stiffness. They are  defined by
\begin{eqnarray}
  v_{N} = \frac{L}{\pi \hbar} \frac{\partial^2 E}{\partial N^2},~~
  v_{\rm s} = \sqrt{\frac{L^2}{mN} \frac{\partial^2 E}{\partial L^2}}.
\end{eqnarray}
In the TLL phase,  the momentum distribution is given by \cite{Cazalilla:2004}
\begin{equation}
  n(k)\simeq A(K) \text{Re}\left[\frac{\Gamma(1/4K + \text{i}kl_{\phi}(T)/2K)}{\Gamma(1-1/4K+\text{i}kl_{\phi}(T)/2K)}\right], \label{Momentum-distribution-LL}
\end{equation}
where $A(K)$ is a $K$-dependent parameter, and $l_{\phi}(T) =\hbar^2\rho_0/(mT)$ is the phase correlation length, see \cite{Cazalilla:2004}.

\begin{figure}[th]
\centerline{\includegraphics[width=0.9\linewidth]{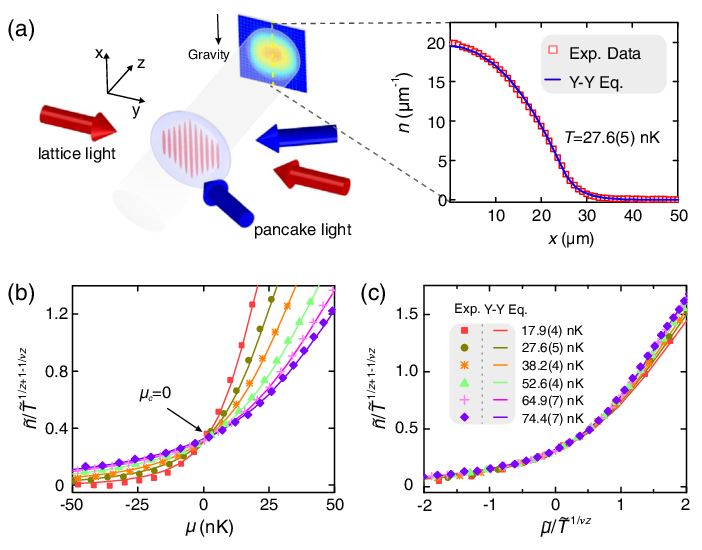}}
\caption{(a) Left: Experimental setup--the quasi-1D trapped Lieb-Liniger gas  consists of an array of tubes  created by a blue-detuned ``pancake" lattice and a red-detuned retro-reflected lattice; Right: the  density profile  was  measured by {\em in situ}  absorption imaging at  the temperature $T=27.6(5)$nK, showing an  agreement  with the prediction of the Yang-Yang thermodynamics equation (\ref{TBA-Bose}).  (b) Quantum scaling of the dimensionless density:  at different temperatures  the densities  intersect at the critical point $\mu_c=0$. The experimental data (symbols) agrees with the theoretical prediction (solid curves) from (\ref{density-scaling}). (c) At different temperatures, the dimensionless  densities against  argument $\tilde{\mu}/\tilde{T}^{1/\nu z}$ collapse into a single curve around $\mu_c=0$. Figure from  \cite{Yang:2017}.}
 \label{fig:Quantum-Criticality-S4}
\end{figure}

In a recent paper  \cite{Yang:2017},  the quantum criticality and Luttinger liquid behaviour of the quasi-1D trapped ultracold ${}^{87}$Rb atoms were measured by {\em in situ} absorption imaging.
In contrast to the usual 3D lattice arrays of 1D tubes, in this experiment,  the density profiles were obtained by {\em in situ} absorption imaging of  single tubes on a single 2D layer.
The density and other thermodynamic scaling laws were obtained at different temperatures and chemical potentials by using a high-resolution microscope.
Fig.~\ref{fig:Quantum-Criticality-S4} shows  the experimental setting and the critical scaling of the density, see  \cite{Yang:2017}.
Here the dimensionless density $\tilde{n}=n/c$ was  the function of dimensionless chemical potential $\tilde{\mu} =\mu/(\hbar^2c^2/2m)$ and dimensionless temperature $\tilde{T}=k_BT/(\hbar^2c^2/2m)$ with $c=-2/a_{1D}$.
The intersection and collapse of density showed in (b) and (c) in the Fig.~\ref{fig:Quantum-Criticality-S4} confirmed the universal scaling law (\ref{density-scaling}).

\begin{figure}[th]
 \begin{center}
 \includegraphics[width=0.9\linewidth]{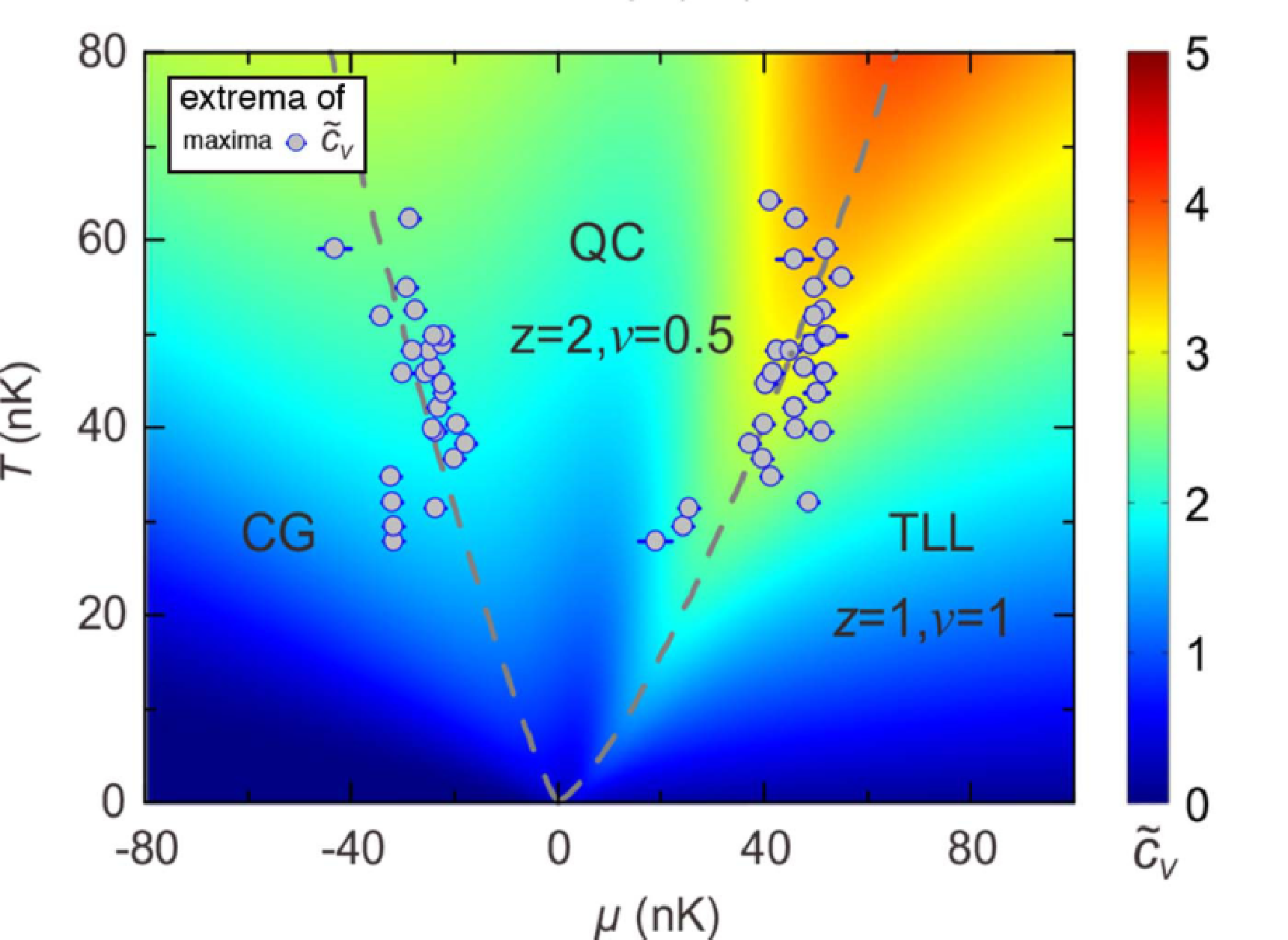}\\
  \includegraphics[width=0.9\linewidth]{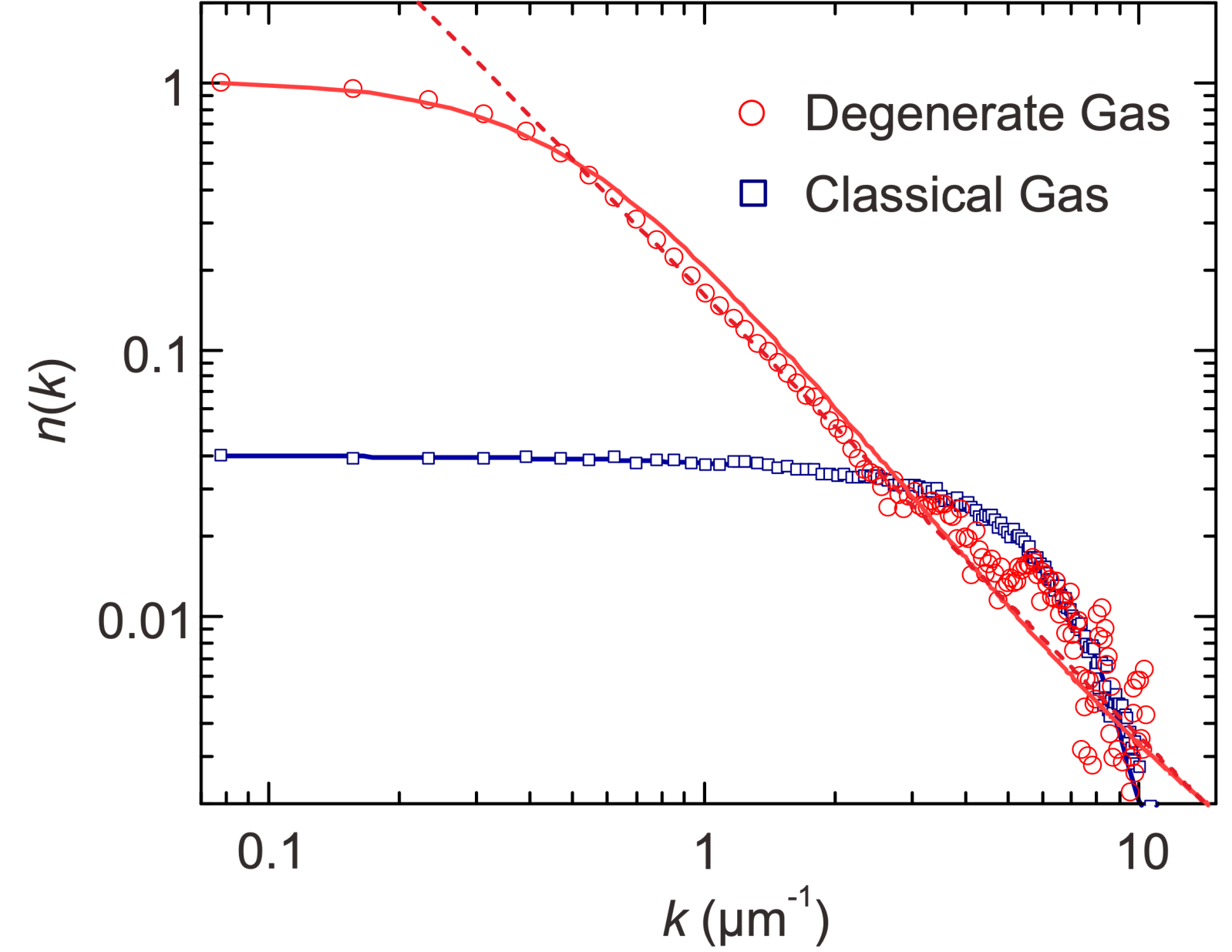}
 \end{center}
 \caption{Upper panel: Contour plot of  the dimensionless specific heat $\tilde{c}_v=c_v/(k_B\,c)$ in $T-\mu$ plane. The filled dots  denoted experimental data of the specific heat peaks that  remarkably mark   two critical crossover temperatures in excellent agreement with the prediction from the TBA equation (\ref{TBA-Bose}).  Lower panel: Momentum distribution of the quasi-1D trapped Lieb-Liniger gas. The red circles and blue squares denote  the experimental momentum distributions at $T=40(1)$ nK and a classical gas at $T=209(1)$ nK , respectively. The red solid curve is  the theoretical prediction  from (\ref{Momentum-distribution-LL})  with  the Luttinger parameter $K = \hbar \pi n/m v_s$.  The red dashed  line  approximately indicates a power-law decay with the slope $-1+1/2K\sim -1.66$, showing a typical TLL correlation. For the ideal Bose gas,  the blue solid curve gives the  Gaussian distribution.
  Figure from  \cite{Yang:2017}. }
 \label{fig:Quantum-Criticality-S5}
 \end{figure}

In  Fig.~\ref{fig:Quantum-Criticality-S5}, the universal phase diagram of criticality was obtained from the measurement of the specific heat in $T-\mu$ plane.
It was found that two crossover branches significantly distinguished  the quantum critical regime from the classical gas (CG) and the TLL phase.
The measured double-peak structure of the specific heat (filled dots) coincides   with the theoretical prediction (dashed lines) based on the TBA equation (\ref{TBA-Bose}), see Fig.~\ref{fig:Quantum-Criticality-S5} (a). In figure (b), the measured momentum distribution at low temperature shows the power-law behavior,  confirming the existence of the TLL.
In particular, the momentum distribution in logarithmic scale had  the asymptotic power-law decay with the slope $-1+1/2K$ at large momenta ($k > 40/l_{\phi}$), see  \cite{Yang:2017}.
Indeed, the experimental result agrees well with the theoretical prediction \cite{Cazalilla:2004}.
A remark made on the determination of the TLL in this Lieb-Liniger gas  \cite{Yang:2017} and the arrays  of Josephson junction \cite{Cedergren:2017}  by T. Giamarchi \cite{Giamarchi:2017}  is very inspiring:\\
``By providing a remarkable experimental demonstration of several so-far-elusive aspects of TLL theory, these two new studies confirm that TLL theory now plays the same key role in 1D systems that Fermi-liquid theory plays in our understanding of 2D and 3D condensed-matter systems. Without a doubt, this research will open new chapters in the TLL field by inspiring studies that examine how other perturbations (coupling between different 1D chains, spin-orbit coupling, and the like) can lead to novel and potentially exotic states in 1D materials."

\subsection{IV.6. Can Quantum Statistics Be Fractional? }

The Bose-Einstein and Fermi-Dirac statistics  play a key role in modern  quantum statistical mechanics.
Now it is well established that they  are not the only possible forms of quantum statistics~\cite{Khare:2005},
for example,  anyonic statistics can occur in excitations in 2D electronic systems, see early studies \cite{Leinaas:1997,Wilczek:1982,Arovas:1984,Laughlin:1988}.
In this regard, a fractional exclusion statistics (FES)  was  introduced   by F D M Haldane in 1991 \cite{Haldane:1991}.
 By counting the Hilbert space dimensionality for available single-particle states in a quantum system,  the number of the available states (holes)  $N_{h,\alpha }$  of species $\alpha $  decreases as particles ($\Delta N _{p,\beta} $ )
 of species $\beta $ are added to the  system via
\begin{equation}
\Delta N_{h,\alpha }=-\sum_\beta g_{\alpha \beta} \Delta N _{p,\beta},
\end{equation}
where the FES parameter $g_{\alpha \beta}$ is independent of the particle numbers $N _{p,\beta} $, but essentially depends on the species $\alpha $ and $ \beta$.  Now it is called   Haldane's mutual FES.  After Haldane's work \cite{Haldane:1991}, Y S Wu \cite{Wu:1994} and other physicists \cite{Ha:1994,Isakov:1994} further formulated the Haldane's FES.
For a non-mutual FES, the parameter  $g_{\alpha \beta}=g\delta _{\alpha, \beta}$.
It is obvious that  the Bose-Einstein and Fermi-Dirac statistics correspond to the non-mutual FES with  $g = 0$ and $1$, respectively.
The FES has been used for the study of macroscopic properties in  several 1D systems, including
Calogero-Sutherland model of particles interacting through a $1/r^2$  potential \cite{Calogero:1969,Sutherland:1969}, the Lieb-Liniger Bose gases \cite{Bernard:1995},  and anyonic gases with delta-function interaction \cite{Kundu:1999,Batchelor:2006}.

On the other hand,  critical phenomena  in nature are ubiquitous. The quantum criticality of the Lieb-Liniger gas which was observed in \cite{Yang:2017} reveals a generic feature of continuous phase transition in interacting many-body systems.
This feature can be commonly found in other systems, like 1D Heisenberg spin chain \cite{HeF:2017} as well as higher dimensional quantum systems \cite{Zhang:2012,Hung:2011}.
Here we ventured to ask if universal nature of quantum criticality can exist beyond the scaling functions and renormalization sense?
In a  recent paper \cite{Zhang:2022}, based on exact solutions, quantum Monte Carlo simulations, and experiments, Zhang {\em et al.} demonstrated that macroscopic physical properties of 1D and 2D Bose gases at quantum criticality can be determined by a simple non-mutual FES.

\begin{figure}[th]
	\includegraphics[width=0.99\linewidth]{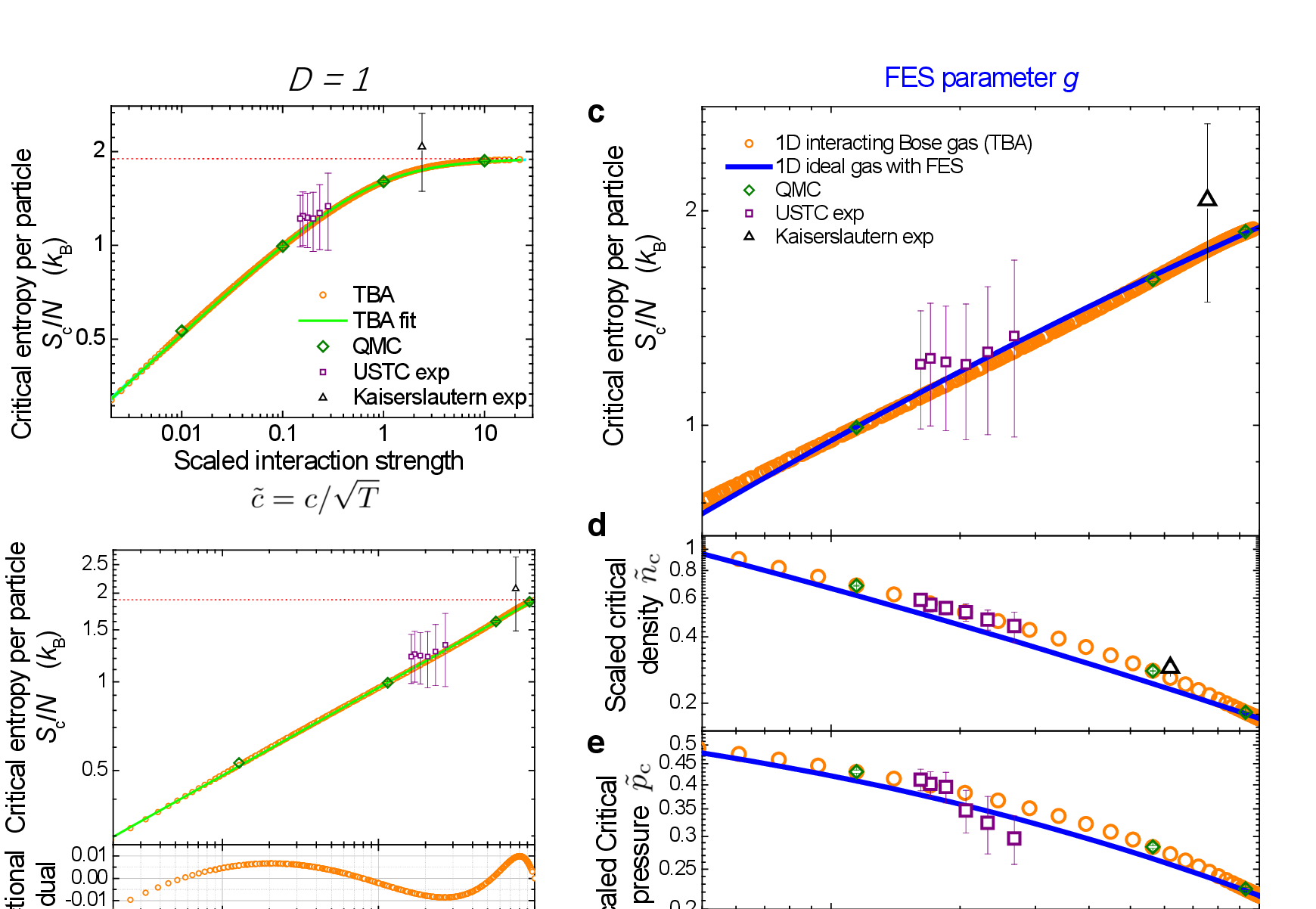}
	\caption{ Evidences for the non-mutual  FES in the 1D Bose gases (\ref{Ham1}) at a quantum critical point $\mu=\mu_c=0$.  (a) Critical entropy per particle $S_{\mathrm{c}}/N$ v.s. the dimensionless interaction strength $\tilde{c}$. The thermodynamical Bethe ansatz solution (circles) is in excellent agreement  with the  QMC computations (diamonds) and experiments (squares and triangles, from Refs.~\cite{Vogler:2013,Yang:2017}. (b)  Power-law scaling of $S_{\mathrm{c}}/N$ with respect to the transformed interaction $\mathcal{C}_{\mathrm{tr}}$. The dotted line denotes the fermionization limit value of $A_{\infty, \mathrm{1D}}$. Here $\beta =0.298 (2)$, $\tilde{c}_1=0.772 (5)$ and $g_{\rm max}=1$, showing a full fermionization in the Tonks-Girardeau limit. 	(c), (d), (e)  Other  thermodynamic observables of  the Lieb-Liniger gas  also agree well with those of ideal particles obeying the  FES. Figure from \cite{Zhang:2022}. }
	\label{fig:FES}
\end{figure}

The FES has a long history of intensive studies, but its experimental realization in physical systems is very rare.
Zhang {\em et al.} \cite{Zhang:2022} found that the non-mutual FES depicts the particle-hole symmetry breaking in 1D and 2D interacting Bose gases at a quantum critical point, see Fig.~\ref{fig:FES}.
In fact, in the vicinity of a quantum critical point,
 the system is strongly correlated with large characteristic lengths in real space.
 Thus in momentum space, the quasi-momentum cells become decoupled into nearly independent cells in which the densities of holes $ \rho_h(k)$ and particles $ \rho_h(k)$ satisfy the following relation
 \begin{eqnarray}
 \rho_h(k)+g\rho(k) =d_{sp},\label{P-H-B}
 \end{eqnarray}
 where $d_{sp}=1/(2\pi)^D$ is the bare dimensionality of states in the unit cell of phase-space  for a $D$-dimensional system.
 This form of particle-hole excitations can be found precisely from the Bethe ansatz
 equations for the Lieb-Liniger model  \cite{Batchelor:2007b}.
 It is obvious  that particle-hole symmetry is broken in Eq. (\ref{P-H-B}).

The physical properties of the system can be calculated through the simple distribution function of ideal particle with a non-mutual FES
\begin{equation}
f(\epsilon) =\frac{1}{w(\epsilon) +g}, \quad w^g(1+w)^{1-g}=\exp\left(\frac{ \epsilon -\mu}{T} \right).
\end{equation}
Here $\mu$ and $T$  denoted  the chemical potential and temperature.
Moreover, the number density and energy density are given by   $n=\int G(\epsilon) f(\epsilon) d\epsilon $ and $e=\int G(\epsilon) f(\epsilon) \epsilon d\epsilon $, where  the density of states per volume is given by $G(\epsilon) = 1/(2\pi\sqrt{\epsilon} )$  in 1D and  $1/4\pi$ for  2D non-relativistic particles.
It is surprising to find that a simple one-to-one correspondence between the interaction strength $c$  in the quantum many-body systems and the statistical parameter $g$ can be given by
\begin{equation}
g=g_{\rm max} {\cal C}_{\rm tr},
\end{equation}
where $g_{\rm max}=1$ is the maximum statistic parameter for the model with an infinitely strong interaction.  The transformed interaction parameter is given by $ {\cal C}_{\rm tr}=\frac{\tilde{c}/\tilde{c}_1}{\tilde{c}/\tilde{c}_1+1}$. Here the
dimensionless interaction strength is given by $\tilde{c}=c/\sqrt{T} $ for 1D Lieb-Liniger gas and $\tilde{c}_1$ is constant.
They found the critical entropy per particle had a power-law dependence on the transformed parameter via $S_c/(Nk_B) =A_\infty {\cal C}^\beta _{\rm tr} $, here  $A_\infty =\frac{3}{2}{\rm Li}_{3/2} (-1) {\rm Li}_{1/2}(-1)$, and $\beta \approx 0.298(1)$ for  the Lieb-Liniger model.
The overall good agreement  with existing experimental data \cite{Vogler:2013,Yang:2017}, thermodynamic Bethe ansatz solution (\ref{TBA-Bose})  and QMC simulation for the thermal properties: critical entropy per particle,  dimensionless density  and pressure was observed in Fig.~\ref{fig:FES}.
It held true for the 2D Bose gas, see a detailed analysis in \cite{Zhang:2022}.
This research raises the question: can  macroscopic thermodynamic properties be  determined by the non-mutual FES  which is described by interaction-induced particle-hole symmetry breaking for the second order quantum phase transition in all dimensions?

\begin{figure}[th]
 \begin{center}
 \includegraphics[width=0.99\linewidth]{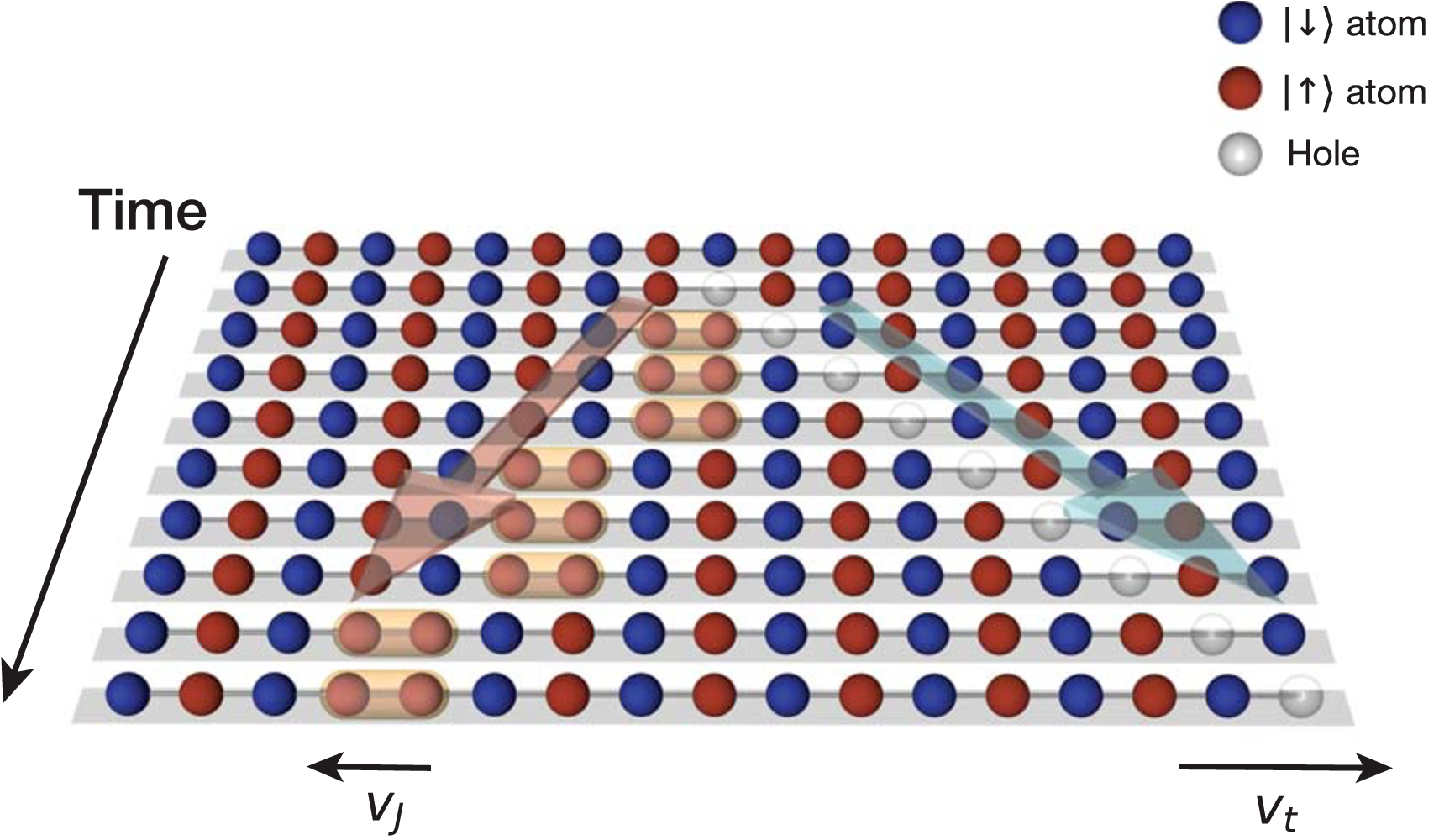}\\
  \includegraphics[width=0.99\linewidth]{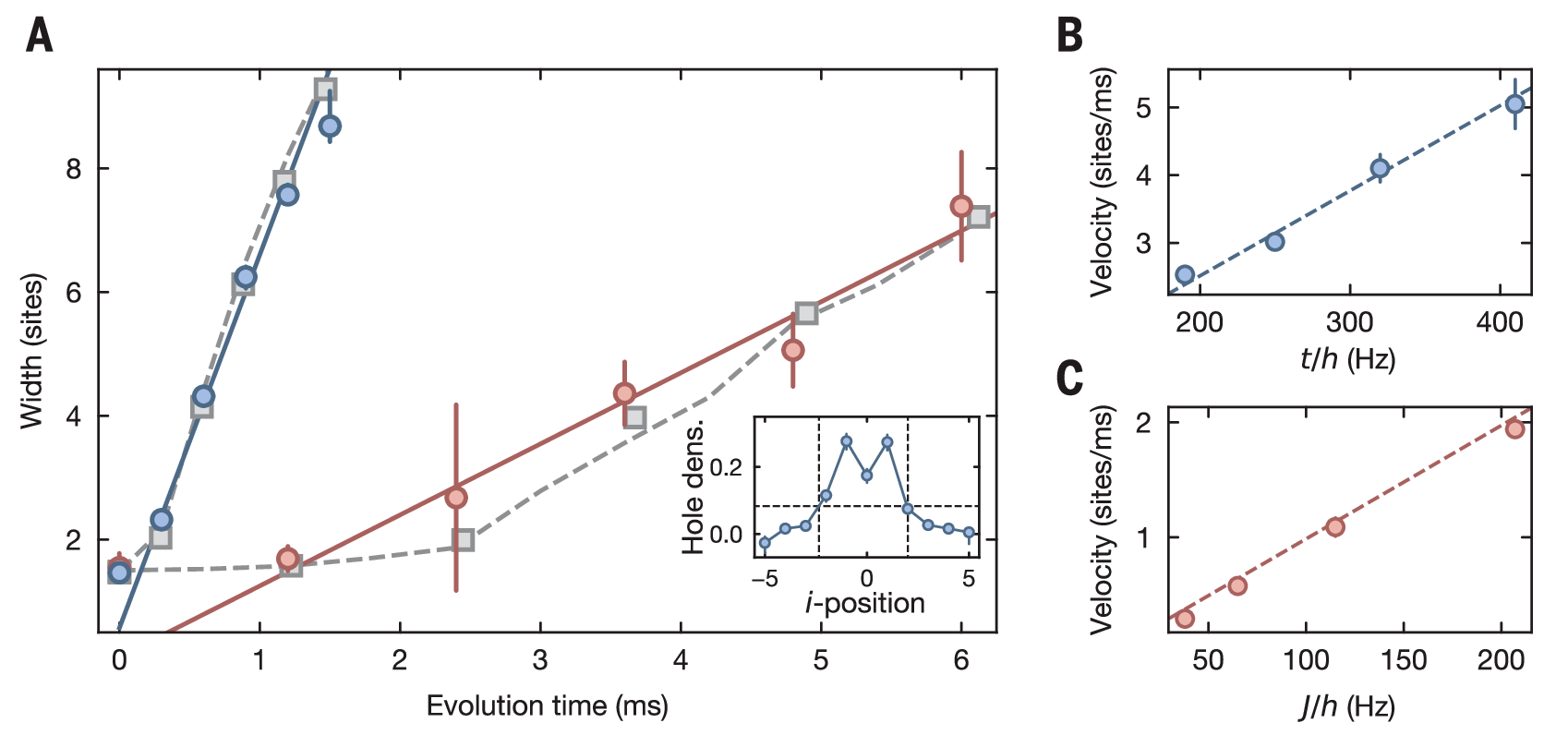}
 \end{center}
 \caption{Upper panel: A schematic demonstration of the separation of a holon and  the spin configuration defect (spinons). The spin-charge decofinement was observed  by monitoring the evolutions of the holons and the spin  after a local quench.
 Lower panel: The quasiparticle sound velocities of holons and spinons.  (A) The width of hole density distribution $\langle n_i^{\rm h} \rangle$ (blue circles)
  and the nearest-neighbour spin correlation $c_{\tilde{i}} (\tilde{x}=1)$ (red circles).
 Their dynamics agrees with the quantum walk of a single particle in density evolution and the antiferromagnetic Heisenberg chain with a light-cone-like propagation, respectively. (B) and (C) show the charge and spin velocities against the  hopping parameter {$t/h $}
  and the effective spin exchange $J/h $ interaction for the 1D strong repulsive Hubbard model. Here $h$ is the Planck's constant.
  Figure from  \cite{Vijayan:2020}. }
 \label{fig:Spin-Charge-separation-S6}
 \end{figure}


\subsection{IV.7. Interacting Fermions in 1D: Spin-Charge Separation}

The TLL  theory describes universal low energy physics of strongly correlated systems of electrons, spins, bosons and fermions in 1D  \cite{Giamarchi-book}.
The TLL theory usually refers to the collective motion of bosons that is significantly different from the free fermion nature of the quasiparticles in higher dimensions.
Such collective nature in 1D many-body systems with internal degrees of freedom results in  remarkable physics, i.e. the so called spin-charge separation--elementary excitations dissolve into  fractional spin  and charge modes with different propagating velocities.
The spin-charge separation as a hallmark of unique 1D many-body phenomenon  has a long history of study in literature.
In solid state materials, evidences for the spin-charge separation were observed in conductance or thermodynamics \cite{Auslaender:2002,Auslaender:2005,Jompol:2009,Segovia:1999,Kim:1996,Kim:2006}.
However, a definite observation of the such phenomenon in term of the Luttinger liquid is still challenging.

Very recently, the  separation of a holon from the two-neighbour spin defect caused by a local  quench dynamics of removing a single particle from an initial Mott state in 1D Hubbard model was studied in   Ref. \cite{Vijayan:2020}.
A single atom was  simultaneously removed from each chain by using an elliptically shaped near-resonant laser.
After such a quench, an  obvious difference in the dynamics of spin and charge degrees of freedom was monitored.
Consequently,  the speed difference between the single hole and the spin configuration  defect caused by  a quench dynamics was verified, see Fig.~\ref{fig:Spin-Charge-separation-S6}.
The dynamical spin-charge separation in the upper panel of Fig.~\ref{fig:Spin-Charge-separation-S6}  was  interpreted by  a quantum walk of single hole in charge sector and
the subsequent evolution of spin configuration induced by  the local quench in the center of the 1D lattices.
Such   dynamical evolutions of independent excitations in charge (holon) and spin (spinon) appear  to give two independent propagating velocities, see the lower panel of the Fig.~\ref{fig:Spin-Charge-separation-S6}.
The dynamics of holons is read off from the specially resolved hole density distribution $\langle n_i^{\rm h} \rangle$ in each Hubbard chain, here $i$ labels  the lattice sizes.
Whereas, the dynamics of spinons was measured by the nearest-neighbour spin correlation
\begin{equation}
c_{\tilde{i}} (\tilde{x}=1)=4\langle\hat{S}^z_{i} \hat{S}^z_{i+1}\rangle-\langle\hat{S}^z_{i}\rangle\langle \hat{S}^z_{i+1}\rangle
\end{equation}
in the squeezed space, see Figure 2 in \cite{Vijayan:2020}.
For a strong interaction, it can be captured by the dynamics of an antiferromagnetic Heisenberg chain.
The spin velocity is proportional to the effective spin exchange coupling constant $J/h$.

The speed difference in the propagation of the single hole and the spinons induced by their quench dynamics provided  an evidence for the spin-charge separation phenomenon.
However,
it  does not mean a confirmation of the spin-charge separation phenomenon described by the Luttinger theory.
The spin-charge separation theory significantly involves collective motion of bosons in charge and spin degrees of freedom, i.e. the elementary low energy excitations of the system dissolve into two Luttinger liquids which solely carry either spin or charge.
Scientifically speaking,
a conclusive  observation of the spin-charge separation predicted by the Luttinger liquid theory  requires:
1) identification of the fractional collective excitation spectra of charge and spin; 2) confirmation of the Luttinger liquid theory in terms of spin and charge correlation functions; 3) determination of spin and charge sound velocities and their Luttinger parameters.

Towards this goal, a recent theoretical study of the Yang-Gaudin model (\ref{Ham2}) discussed the microscopic origin of the spin-charge separation and gave a proposal for experimental measurement of this phenomenon  \cite{He:2020}.
Following this theoretical scheme, R. Hulet group from Rice University with his theoretical collaborators has observed the independent spin and charge density waves in this model with tunable repulsive interaction \cite{Senaratne:2021}.
In this experiment, the spin and charge density waves which were  encoded by their dynamical correlation functions not only confirm the nature of the Luttinger liquids of spin and charge, but also the nonlinear effects due to the band curvature in charge excitation and backward scattering spin excitation.

Before introducing  a new  experimental observation of spin-charge separation in the Yang-Gaudin model \cite{Senaratne:2021}, here we first recall several concepts of the TLL theory which is used to describe the low energy physics of the Yang-Gaudin model.
Following the notation in the book  \cite{Giamarchi-book}, the effective Hamiltonian for the linear charge and spin excitations reads
\begin{eqnarray}
&&H_{LL}=\frac{1}{2\pi}\int dx\left[ u_\rho K_\rho(\pi\Pi_\rho (x))^2+\frac{u_\rho}{ K_\rho}(\nabla\phi_\rho (x))^2\right]  \nonumber \\
&&+
\frac{1}{2\pi}\int dx\left[ u_\sigma K_\sigma(\pi\Pi_\sigma (x))^2+\frac{u_\sigma}{ K_\sigma}(\nabla\phi_\sigma (x))^2\right],
\end{eqnarray}
where $\phi_\rho$ and $\phi_\sigma$ represent the two independent bosonic  fields.
In the above equations,  the fields $ \phi_{\nu}$ and its canonically conjugate momentum $\Pi_{\nu}$ obeys the standard Bose communication relations , i.e.
$\left[ \phi_{\nu},\Pi_{\mu} \right]=\mathrm{i} \delta_{\nu\mu}\delta(x-y)$ with $\nu,\mu=c,\sigma$.
The $\Pi_{\mu}=\partial_x \theta_{\nu} (x)/\pi$  stands for the variation of a  bosonic field $\theta_\nu(x)$.
This effective Hamiltonian characterizes the long distance asymptotic decay of correlation function for the 1D Fermi gas.
The coefficients  for different processes are given phenomenologically in \cite{Giamarchi-book}.
The susceptibilities of charge and spin are defined by
\begin{eqnarray}
		\chi^{\rho\rho}(q,\omega)=-i\int dt\,dxe^{i(\omega t-qx)}\theta(t)\langle [\rho(x,t),\rho(0,0)]\rangle,\\
		\chi^{\sigma\sigma}(q,\omega)=-i\int dt\,dxe^{i(\omega t-qx)}\theta(t)\langle [\sigma(x,t),\sigma(0,0)]\rangle,
\end{eqnarray}
where the charge density and spin density are defined by
\begin{eqnarray}
		\rho(r)&=&\frac{1}{\sqrt{2}}\left[\rho_\uparrow(r)+\rho_\downarrow(r) \right],\nonumber  \\
		\sigma(r)&=&\frac{1}{\sqrt{2}}\left[\rho_\uparrow(r)-\rho_\downarrow(r) \right]. \nonumber
\end{eqnarray}
From the bosonization approach, if $q\ll k_F$, the two dynamic response functions can be obtained by only considering the contributions of the variations $\nabla\phi_\rho(r)$ and $\nabla\phi_\sigma(r)$ in charge and spin densities, namely
\begin{eqnarray}
		\chi^{\rho\rho}(q,\omega)&=&\frac{q^2K_\rho u_\rho}{\pi((u_\rho q)^2-(\omega+i\epsilon )^2)}, \\
		\chi^{\sigma\sigma}(q,\omega)&=&\frac{q^2K_\sigma u_\sigma}{\pi((u_\rho q)^2-(\omega+i\epsilon )^2)}.
\end{eqnarray}
Here $K_{\rho,\sigma}$ are the Luttinger parameter for charge and spin, respectively.

However, the interplay between the spin and charge degrees of freedom leads to subtle difference in effects of  band curvatures in these fractionalized spin and charge excitations. Such nonlinearity dispersions in spin and charge degrees of freedom require the newly developed  nonlinear TLL  theory \cite{Imambekov:2009,Imambekov:2012}.
Whereas for charge sector in the low temperature regime, i.e.,  $T\ll T_F=mv_F^2/2$ and $q \ll k_F$,    the broadening of the charge DSF associated the dispersion  curvature $q^2$ terms  can be well approximated by the imaginary part of the density-density correlation function for free fermions \cite{Pereira:2010,Cherny:2006}.
The charge DSF of 1D noninteracting homogeneous  Fermi gas is given by \cite{Cherny:2006}
\begin{eqnarray}
\label{LDSorigin}
S(q,\omega)=\frac{{\rm Im} \chi (q, \omega,k_F,T,N)}{\pi (1-{\rm e}^{-\beta \hbar \omega})},
\end{eqnarray}
where $\chi$ is the dynamic polarizability, its imaginary part can be written by
\begin{eqnarray}
\label{finiteTDSF}
{\rm Im} \chi (q, \omega,k_{F},T,N)=\frac{N  m^*}{2 \hbar^2 q k_F} \pi (n_{q_{-}}-n_{q_{+}})
\end{eqnarray}
with
\begin{eqnarray}
\label{fermidistribution}
q_{\pm}=\frac{\omega m^*}{\hbar q}\pm \frac{q}{2},\quad n_q=\frac{1}{{\rm e}^{\beta(\varepsilon -\mu)}+1},\quad \varepsilon=\frac{\hbar^2 q^2}{2 m^*}.
\end{eqnarray}
Here $m^*$ is the effective mass of the interacting fermions \cite{He:2020}.

\begin{figure}[th]

 \includegraphics[width=0.66\linewidth]{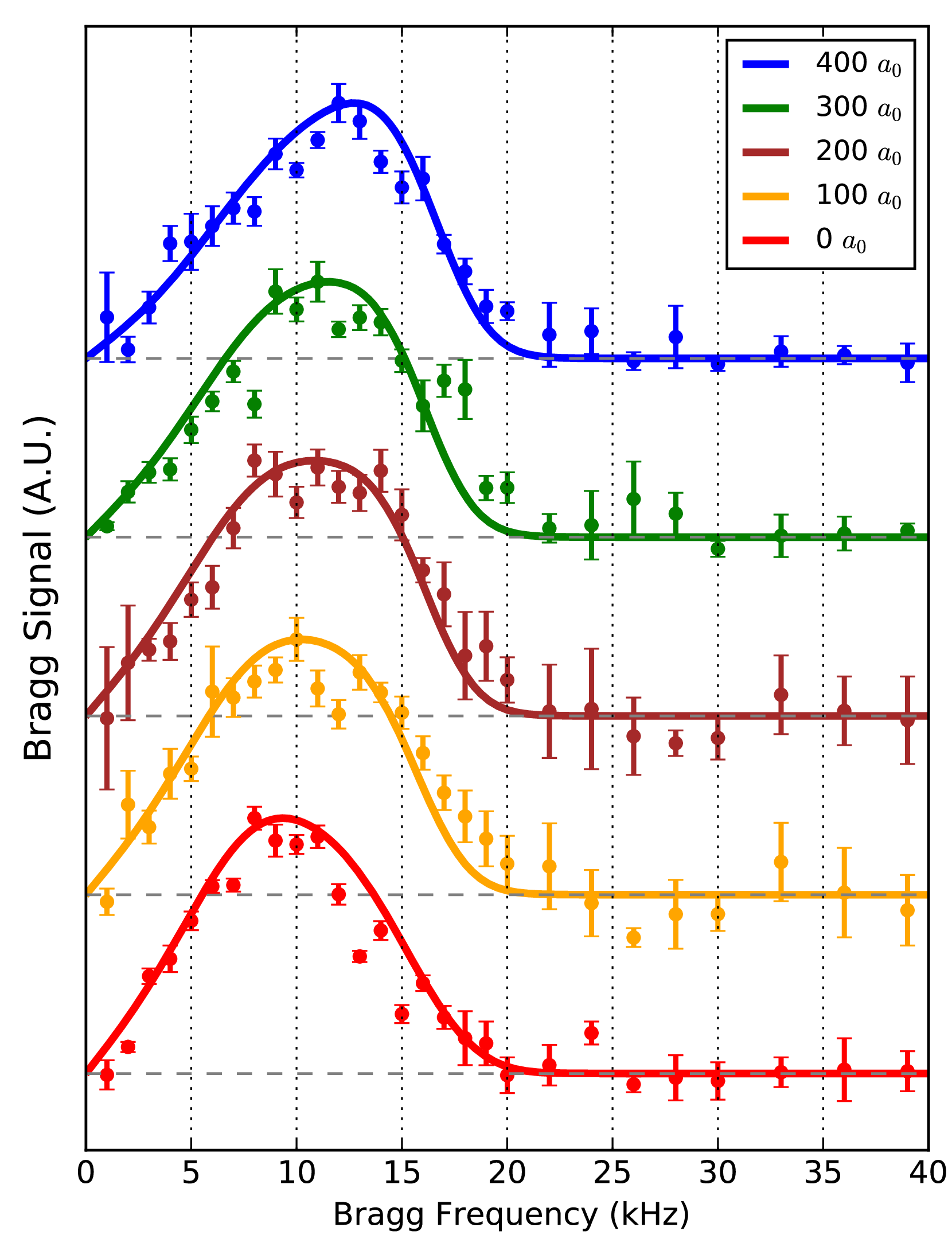}

  \includegraphics[width=0.7\linewidth]{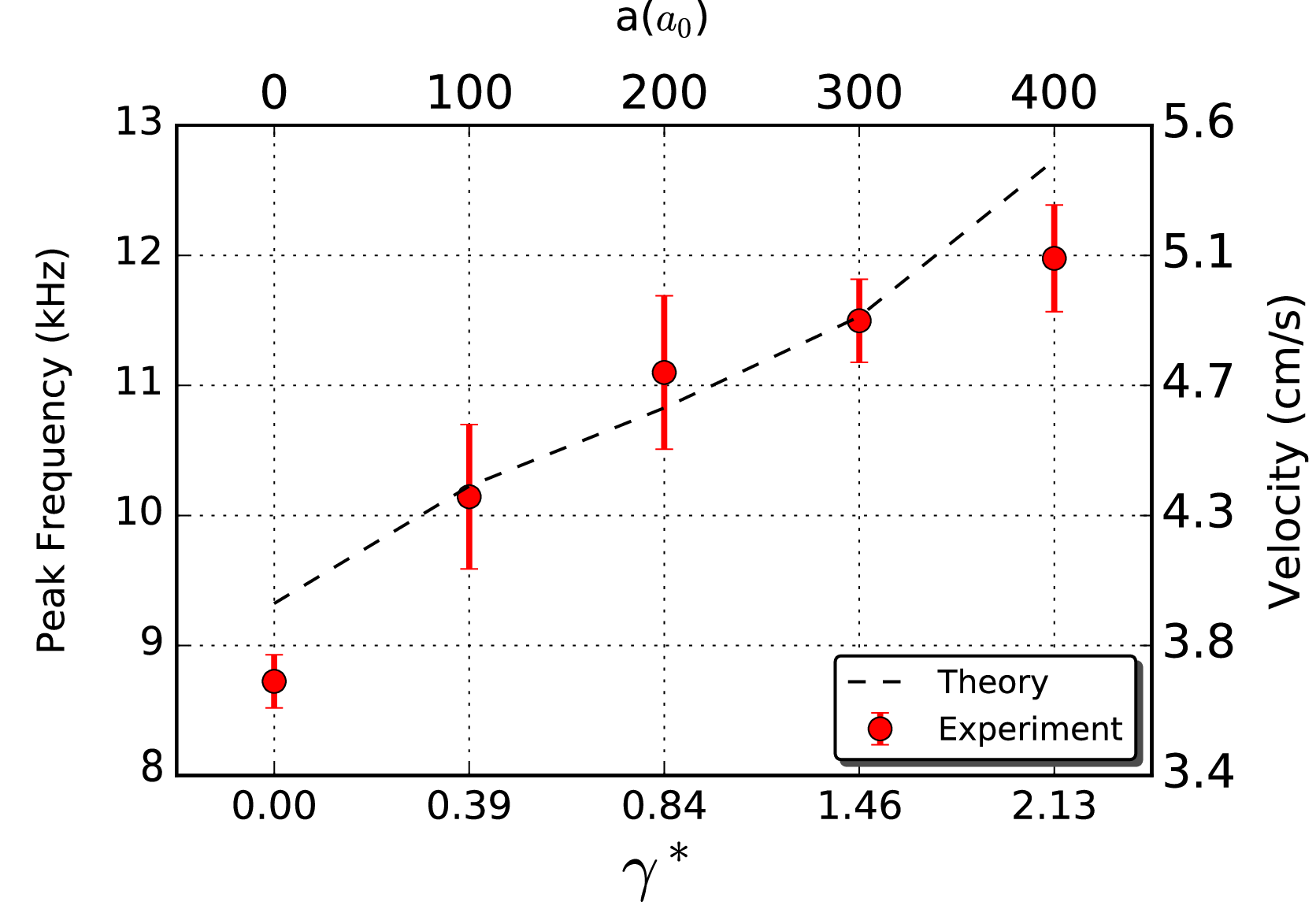}

 \caption{Upper panel:
 Bragg spectral signal $S_{C}(q,\omega)$ v.s. detuning frequency $\omega$ for the quasi-1D trapped Yang-Gaudin model (\ref{Ham2})). Normalized Bragg data, which are  related to the charge DSF  $S_C(q,\omega)$ (symbols),   are in good agreement with theoretical simulation from  Eq. (\ref{LDSorigin}) (solid curves) for the range of 3D scattering length $a$ from 0 to 400 $a_0$. In theoretical simulation  a global temperature $T = 200$ nK and the local density approximation were used.
 Lower  panel: The peak frequency (left vertical axis) and sound velocity (right vertical axis) v.s. the scattering length for $a= 0, \, 100,\, 200,\, 300,\, 400$ $a_0$. The corresponding values of the effective dimensionless interaction $\gamma*$ in the center of the tube with the most probable number were shown in  horizontal axis. The black dashed line is the Bethe ansatz  theoretical value for each interaction strength.
  Figure from  \cite{YangTL:2018}. }
 \label{fig:Charge-DSF}
 \end{figure}

\begin{figure}[th]

 \includegraphics[width=0.66\linewidth]{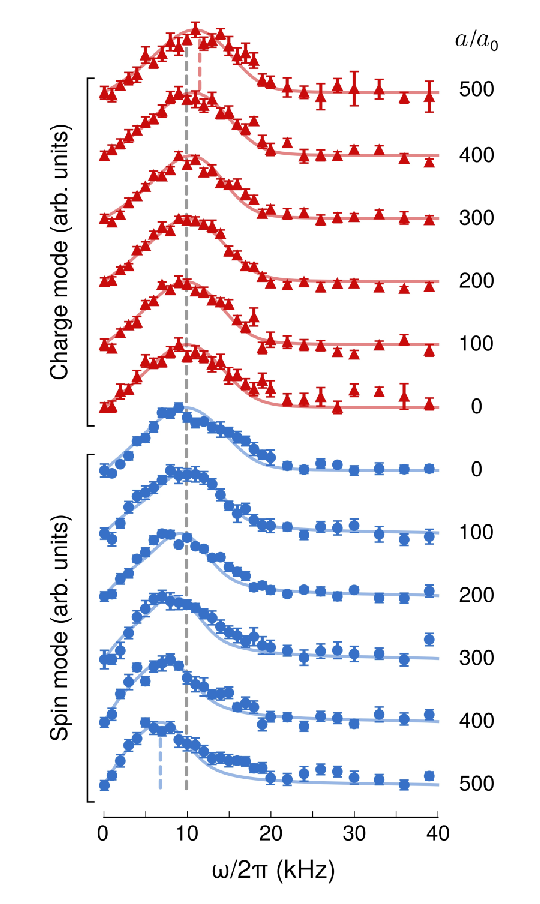}
 \includegraphics[width=0.69\linewidth]{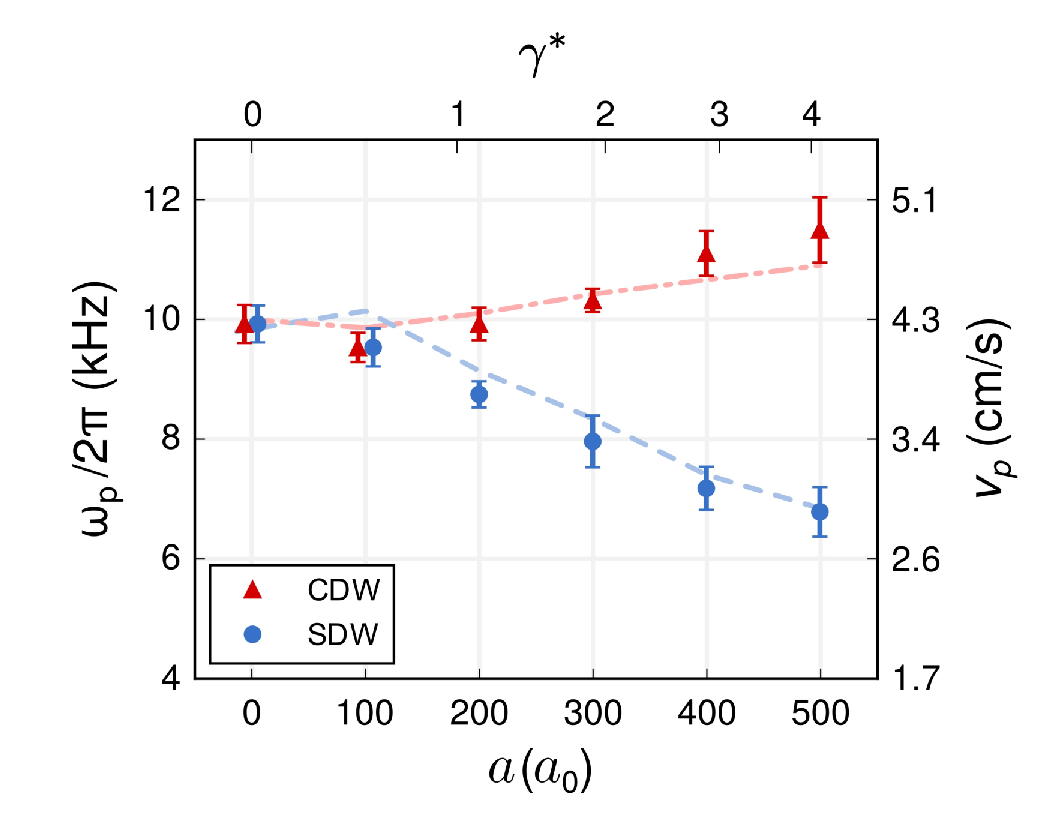}
 \caption{Upper panel:
 Bragg spectra  v.s. detuning frequency $\omega/2\pi $ for the charge and spin density waves. The normalized Bragg signals $S_C(q,\omega)$ (red triangles)(symbols) and $S_S(q,\omega)$ (blue circles)  are in good agreement with theoretical simulation (solid curves)   for the range of 3D scattering length $a$ from 0 to 500 $a_0$ and for a global temperature $T$ = 250 nK. The long  vertical dashed line shows the extracted peak frequency $\omega_p$ for the non-interacting case (gray), and the short vertical dashed lines indicate the strongest probed interactions for the spin- and the charge-modes, respectively.
 Lower panel: Spin-charge separation. The peaks of measured Bragg spectra for charge (red triangles) and spin (blue circles) agree well with theoretical velocities (red and blue dashed lines)  for the range of 3D scattering length $a$ ranging from 0 to 500 $a_0$.
  Figure from  \cite{Senaratne:2021}. }
 \label{fig:Spin-Charge-separation-S8}
 \end{figure}

In the paper \cite{Pereira:2010}, R G Pereira and E Sela derived the nonlinear effects in term of the spin-charge coupling.
It turns out that the backward scattering term $H_{\rm backward}=\frac{2g_1}{(2\pi \alpha)^2}\int dx \cos (\sqrt{8}\phi_{\sigma})$ in spin sector is the next leading order contribution to the low energy excitations. The parameter $\alpha$  is a short-distance cutoff.
The effective Hamiltonian for the spin excitation in low energy limit can be given by
\begin{eqnarray}
H_{\rm \sigma}&=&
\frac{1}{2\pi}\int dx\left[ u_\sigma K_\sigma(\pi\Pi_\sigma (x))^2+\frac{u_\sigma}{ K_\sigma}(\nabla\phi_\sigma (x))^2\right]\nonumber\\
&&+ \frac{2g_1}{(2\pi \alpha)^2}\int dx \cos (\sqrt{8}\phi_{\sigma}).
\end{eqnarray}
The spin DSF,  at zero temperature, is a $\delta$-function peak at $\omega = v_s q$.
However, at low temperatures, one does  not have a similar approximation of the DSF in term of noninteracting spinless fermions.
Nevertheless, by considering the $g_1$ interaction, the propagator of the dressed spin bosons is given by \cite{Pereira:2010}
\begin{eqnarray}
\widetilde{S}(q,i\omega)=\frac{1}{4\pi}\frac{q}{i\omega-v_sq-\sum(q,i\omega,T)},\label{Spin-propagator}
\end{eqnarray}
here $\sum (q,i\omega,T)$ is the self-energy of spin bosons.
It is straightforward to obtain the spin DSF
\begin{eqnarray}
\text{Im}\chi=\frac{2K_s\tau_s(T)}{\pi[\tau_s(T)(w-v_sq)]^2+\pi}.\label{Spin-DSF}
\end{eqnarray}
We assumed that the real part of the retarded self-energy is zero, whereas $\tau_s$ is given in  term of the imaginary part of $\sum(q,i\omega,T)$ 
\begin{eqnarray}
{\rm Im}\sum_q=-\frac{1}{\tau_s(T)}=-\frac{\pi }{2}\left[g(T) \right]^2k_BT,
\end{eqnarray}
where the  renormalization coupling constant $g(T)$  can be determined by \cite{Vijayan:2020}
\begin{eqnarray}
g(T)\approx\frac{g}{1+g\ln(T_F/T)}
\end{eqnarray}
with  $T_{F}\sim mv_{F}^{2}$ and g=$g_1/(\pi v_s)$, here $g_1=c$ is the interaction strength of the model.

Using Bragg beams to excite charge and spin excitations with the momentum transfer to the system
from the Bragg beams is given by
\begin{equation}
P(q,\omega) \propto \left(\frac{1}{\Delta^2_\uparrow}+\frac{1}{\Delta^2_\downarrow}\right)S_{\uparrow\uparrow}+\frac{2}{\Delta_\uparrow\Delta_\downarrow}S_{\uparrow\downarrow},
\label{eq:pD}
\end{equation}
where $\Delta_\sigma$ is the relative detuning of the Bragg beam with respect to each spin state.
For charge excitation in experimental setting with $\Delta_{\uparrow} \approx \Delta_{\downarrow} \gg \Delta_{\uparrow\downarrow}$, here $\Delta_{\uparrow\downarrow}$ is the  splitting energy between the two spin states, then the momentum transfer  $P(q,\omega) \propto S_C(q,\omega)$ for the change density wave. Whereas  for spin excitations, $\Delta_{\uparrow} = - \Delta_{\downarrow} = |\Delta_{\uparrow\downarrow}|/2$, then  the momentum transfer  $P(q,\omega) \propto S_S(q,\omega)$. In the above equations, the charge- and spin-density DSF are given by
\begin{equation}
S_{C,S}(q,\omega) \equiv 2\left[S_{\uparrow\uparrow}(q,\omega) \pm S_{\uparrow\downarrow}(q,\omega) \right].
\label{eq:s}
\end{equation}
The two independent DSFs  $S_{\uparrow\uparrow}$ and $S_{\uparrow\downarrow}$ can be calculated from the dynamic polarizability discussed above.
At  finite temperatures, the momentum transfer is modified accordingly as $P(q,\omega) \propto S(q, \omega)-S(-q, -\omega)=S(q, \omega)(1-\mathrm{exp}(-\hbar\omega/k_\mathrm{B} T))$, see \cite{Senaratne:2021}.

The excitation spectrum of the charge density wave of the Yang-Gaudin model (\ref{Ham2})  was experimentally observed   for various  interaction strengths by R. Hulet group at Rice university  \cite{YangTL:2018}.
In this paper they directly measured the charge DSF for a fixed momentum $q\simeq 0.2 K_F$ in the ensemble of ${}^6$Li  atoms in the two energetically  hyperfine sublevels  $|1\rangle$ and $|2\rangle $ ($|F=1,M_F=\pm1/2$), see Fig~\ref{fig:Charge-DSF}.
They were for first time to demonstrate experimentally  that   the charge excitation in the 1D interacting fermions can be approximately described by the DSF of free fermions (\ref{LDSorigin}).
The agreement between the Bragg signal and theoretical simulation can be  seen in  Fig~\ref{fig:Charge-DSF}.
They further showed that the velocities read off from the  Bragg peak frequency  agreed with the results obtained from Bethe ansatz
\cite{Guan:2013}.
 The significance of this observation  is the manifestation of the Luttinger liquid nature in the collective charge excitation.

In contrast to the Bragg spectroscopy  for the  charge excitation, the measurements of the spin Bragg spectra requires that the Bragg photons must be much closer detuned from resonance with the chosen excited state than for the charge-mode measurement. In the recent  experiment \cite{Senaratne:2021}, a spin-balanced mixture of $^6$Li atoms  was realized in an anisotropic optical trap.
In this work, by  applying a pair of Bragg beams on the sample in a 200 $\mu$s pulse, the spin and charge DSFs were independently measured for the quasi-1D trapped Yang-Gaudin Fermi gases, see Fig.~\ref{fig:Spin-Charge-separation-S8}.
By using the pseudo-spin-$1/2 $ states $|1\rangle $  and $|3\rangle $, the detuning from the excited state was reduced by about a factor of two, and the rate of spontaneous emission is correspondingly reduced.  More significantly, however, the rate of spontaneous emission was reduced further by using the narrow $2S-3P_{3/2}$ (UV) transition for the Bragg spectroscopy, rather than the usual $2S-2P_{3/2}$ (red) transition.  Thus the decay linewidth of the $3P_{3/2}$ state is approximately $8$ times narrower than the $2P_{3/2}$ state \cite{Duarte:2011,Cherny:2006}, resulting in a total reduction of spontaneous emission for a given Bragg signal by a factor of $~16$, in comparison to the previous experiment \cite{YangTL:2018}.

 To measure the spin Bragg spectra, they chose the states $|1\rangle $  and $|3\rangle $  as the  pseudo-spin-$1/2 $ states in order to
decrease the detuning of the Bragg beams from resonance to reduce the
spontaneous emission rate.

The Fig.~\ref{fig:Spin-Charge-separation-S8} shows the measured (symbols) and calculated (solid lines) Bragg spectra for both spin and charge modes in the range of the interactions with  the 3D scattering length $a$ from 0 to 500 $a_0$, here $a_0$ is the Bohr radius.
The Bragg signal was compared with the calculated charge and spin spectra by  invoking the local density approximation, showing excellent agreement between experimental data and theoretical simulations of the charge and spin DSFs from the Eqs.
(\ref{finiteTDSF}) and (\ref{Spin-DSF}).
This agreement significantly confirms the linear Luttinger liquid dynamical correlations   with the nonlinear contributions from the curvature effect  in charge degree of freedom and backward scattering effect in spin degree of freedom.
Moreover, the frequency at which the Bragg signal reaches a maximum, $\omega_\mathrm{p}$, corresponds to the values of the sound  velocities of charge and spin via   $v_\mathrm{p}  = \omega_\mathrm{p}/q$, in the ensemble of 1D tubes.  The lower panel of the Fig.~\ref{fig:Spin-Charge-separation-S8} shows agreement between the velocities extracted from the peak frequencies  (symbols) and theoretical ones (dashed lines) obtained from the Bethe ansatz solution.
The hallmark of the 1D many-body physics--spin-charge separation has been elegantly confirmed in this work.
A more detailed study of such novel phenomenon, see  \cite{He:2020,Senaratne:2021}, also the pedagogical book \cite{Giamarchi-book}.

Experimental measurements of phase diagram and pairing of two-component ultracold ${}^{6}Li$  atoms trapped in an array of 1D tubes were reported in \cite{Liao:2010,Revelle:2016}. The trapped fermionic atomic gas was demonstrated a spin population imbalance caused by a difference in the numbers of spin-up and spin-down atoms. Such a novel phase diagram was further investigated by the 1D to 3D crossover of the Yang-Gaudin model with tunable interaction strengths  \cite{Revelle:2016}.

\begin{figure}[th]

 \includegraphics[width=0.89\linewidth]{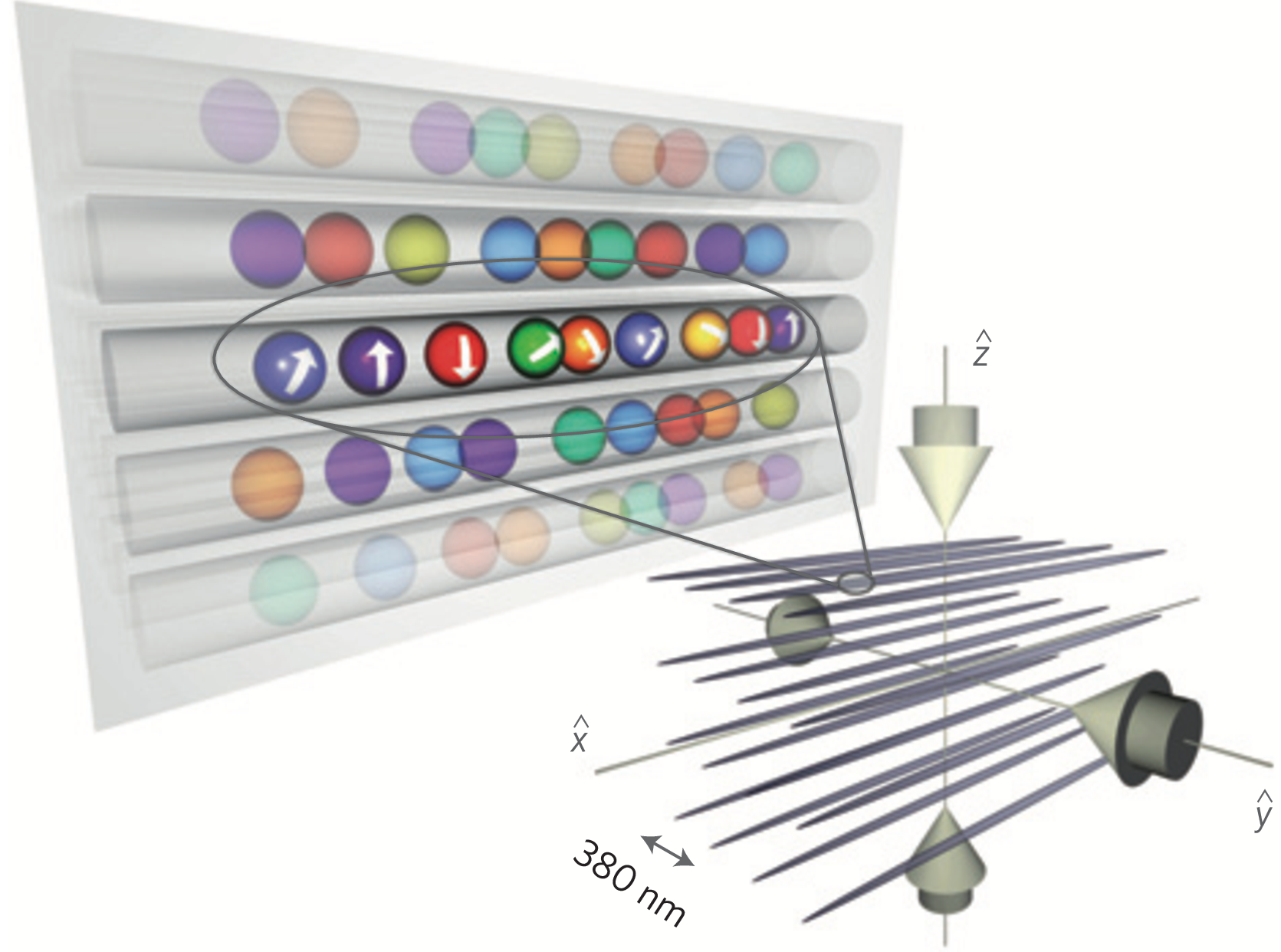}\\

 \vspace{0.3cm}

 \includegraphics[width=0.89\linewidth]{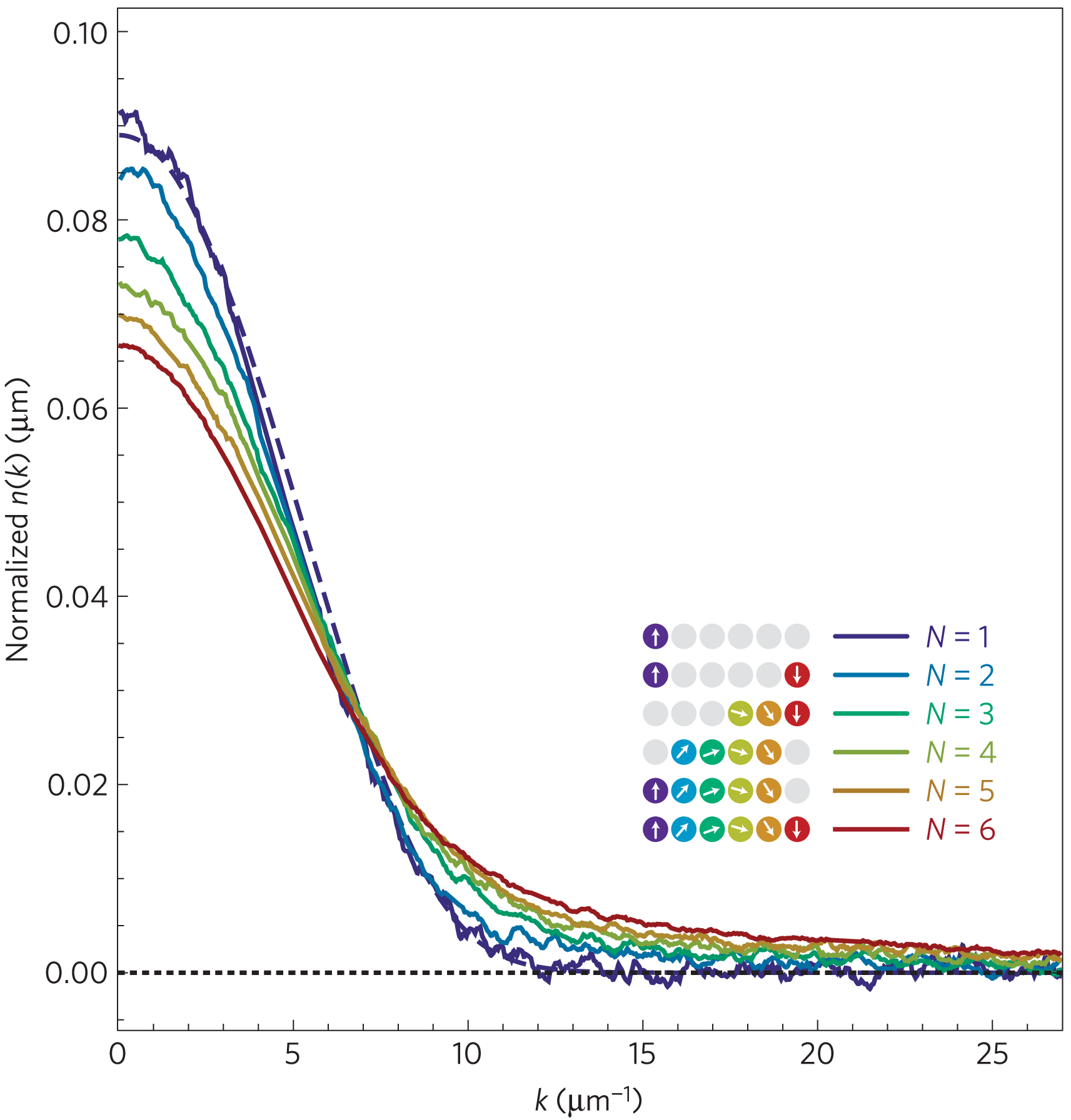}
 \caption{Upper panel: Experimental realization of 1D ultracold Fermi gases of ${}^{173}$Yb atoms  with tunable $SU(w)$ symmetry. The ensemble of 2D lattice arrays of 1D wires is  set for the degree changes from $w=1$ to $w=6$.
 Lower panel:  The momentum distribution $n(k)$  of the 1D Fermi gas with different spin degrees of freedom. The curves in different color shows degree changes from $w=1$ to $w=6$. Here the number of particles in each spin state is the same. The dashed line is the theoretical result of the momentum distribution  for the free Fermi gas.
  Figure from  \cite{Pagano:2014}. }
 \label{fig:SUN-momentum}
 \end{figure}

\subsection{IV.8. Multicomponent Interacting Fermions With Tunable Spin}

 Fermionic alkaline-earth atoms exhibit an exact $SU(w)$  spin symmetry \cite{Cazalilla:2009,Gorshkov:2010}.
 This discovery has led to rich experimental developments \cite{Taie:2010,XZhang2014S,Cappellini2014PRL, Scazza2014NP}, including  Kondo spin-exchange physics \cite{Nakagawa:2015,Zhang2015}, $SU(w)$ fermionic Hubbard model \cite{Ozawa:2018,Taie:2020} and  Tomonaga-Luttinger liquid (TLL) \cite{Pagano:2014} etc.
Such multicomponent interacting fermions provide an ideal platform to simulate unusual symmetries in nature.
The interplay between the symmetries and interaction leads to striking features of quantum dynamics, thermodynamics as well as magnetocaloric-like effects see recent new developments \cite{Taie:2012,Hofrichter:2016,Goban:2018,Song:2020}.
The ultracold alkaline-earth atomic gases also provide promising applications in quantum precision measurements.

In term of quantum integrability,  non-diffraction in the many-particle scattering process displays subtle fractional spin, charge excitations  and criticality in the integrable models of interacting fermions with higher symmetries, see review  \cite{Guan:2013}.
In this scenario, the charge excitation  mode of fermionic atoms confined to quasi-1D tubes has been previously measured for the 1D $SU(w)$ Fermi gas with a fixed interaction \cite{Pagano:2014}.
In this experiment, the authors realized an ensemble of 1D wires of repulsive $SU(w)$ Fermi gases of ${}^{173}$Yb ultracold atoms with tunable degree $w$ from $w=1$ to $w=6$. 
The pure nuclear spin $I=5/2$ results in spin independent interaction such that the system has up to $SU(6)$ symmetry, see the upper panel in Fig.~\ref{fig:SUN-momentum}.
The authors in  \cite{Pagano:2014} demonstrated the role of the spin degrees of freedom by measuring the momentum distribution $n(k)$ for $w=1$ to $w=6$.
They observed that the broadening of the momentum distribution with a reduction of the weight at low momentum and slower decay at the large momentum tails when the number of spin components increases, see the lower panel in Fig.~\ref{fig:SUN-momentum}.
To understand this feature, we see that the ground state energies for the 1D $SU(w)$ Fermi gases with weak and strong repulsions \cite{Guan:2012b}
\begin{eqnarray}
E& \approx &\frac{\pi^{2}n^{3}}{3w^{2}} +c(w-1)n^2/w +O(c^{2}), \label{E-R1}\\
 E&\approx & \frac{n^3\pi^{2}}{3}\left\{1
-\frac{4 Z_{1}}{\gamma }+
\frac{12 Z_{1}^{2}}{\gamma^2 }- \frac{32 }{\gamma^3}\left(  Z_{1}^{3}-
\frac{Z_{3}\pi^{2}}{15}\right) \right\}\label{E-R2}
\end{eqnarray}
have less Fermi kinetic energy when the spin degrees increase.
In the above equation $Z_1=- \frac{1}{w} \left[\psi(\frac{1}{w})+C\right]$
 and $Z_3= w^{-3}\left[ \zeta(3,\frac{1}{w})-\zeta(3)\right]$.
Here $\zeta(z,q)$ and $\zeta(z)$ are the Riemann $\zeta$ functions, $\psi(p)$ denotes the Euler $\psi$ function, $C$ denotes the Euler constant.
Therefore, the Fermi gas with larger internal degrees of freedom has less Fermi pressure.
Consequently, the trapped $SU(w)$ Fermi gases in the same axial trapping potential  would be more spreading in momentum distribution with increasing the component $w$.

In work  \cite{Pagano:2014}, by measuring the square ratio of the  breathing frequency $\omega_B$ to the trapped axial frequency $\omega_x$, i.e. $\beta =(\omega_B/\omega_x)^2$,   the authors further demonstrated an important feature that  the ground state energy of the 1D multi-component Fermi gases reduces to the one of the spinless Bose gas.
The ratio square $\beta$ reflects this feature of the  ground state energy, see \cite{Liu:2014}.
Their experimental observation  can be understood by the calculation of the parameter $\beta$ via  the ground state energy of the model.
The ground state energy given by  (\ref{E-R1}) and (\ref{E-R2}) reduce to the one of the Lieb-Liniger gas when we take $w\to \infty$.
For a strong repulsion and in this limit $w\to \infty$,  it was observed \cite{Jiang:2015} that the spin velocity $ v_{\rm s}$ vanishes while the charge velocity $v_{\rm c}$  turns to the one for the Lieb-Liniger Bose gas, namely,
\begin{eqnarray}
  v_{\rm s}&=&\frac{4\pi^3n^2}{3wc}(1-6 g_0 n/c),\nonumber\\
 v_{\rm c}&=&2\pi n(1-4 g_0 n/c).
\end{eqnarray}
Here $g_0=\frac{1}{w}\left(\zeta(\frac{1}{w}-\zeta (0) \right)$.
This clearly  shows  a strong  suppression of the Fermi exclusion statistics by the commutativity feature among the $w$-component fermions with different spin states, which was for the first time found by Yang and You in \cite{Yang-You:2011} and confirmed by Guan {\em et al.} \cite{Guan:2012b}.
This feature was recently confirmed in higher dimensional ultracold atoms \cite{Song:2020}.
Although they also measured the charge excitation spectrum for different spin states by Bragg spectroscopy tuning, observation of the charge and spin velocities as well as the spin-charge separation in the 1D multicomponent Fermi gases still remains challenging.

In addition, the antiferromagnetic correlations in the 1D, 2D and 3D $SU(w)$ Hubbard models were studied in \cite{Taie:2020,Hart:2015,Singha:2011,Schneider:2012,Cheuk:2016,Parsons:2016,Fujita:2017}.
Extending the $SU(2)$ symmetry in the Mott insulators of the Hubbard model to such higher $SU(w)$ symmetries could lead to richer magnetic ordering and magnetism than that of the usual electronic correlated systems.
The 1D spinor bosonic Hubbard models have provided  novel realizations of spin wave quasiparticles (magnons) \cite{Fukuhara:2013a,Fukuhara:2013b}, and antiferromagnetic correlations \cite{Sun:2021}.
Nevertheless, the multicomponent Fermi and Bose gases with tunable mathematical symmetries open to investigate  fundamental many-body spin dynamics and magnetism for future quantum technology, for example, fractional heavy  spinons and magnons in the 1D $SU(w)$ Fermi and Bose gases have not observed in experiment.

\begin{figure}[th]

 \includegraphics[width=1.0\linewidth]{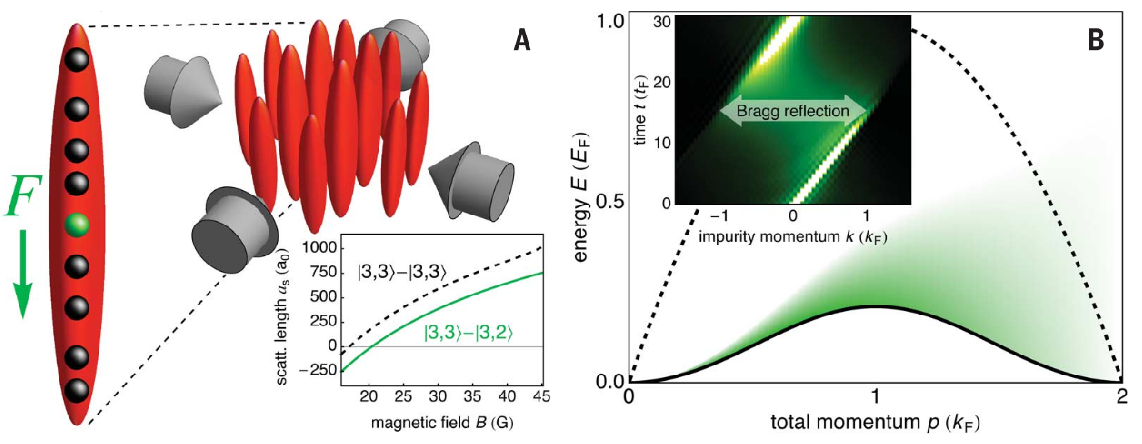}
 \caption{(A) An experimental ensemble of 1D Bose gas of Cesium atoms of the state $|F,m_F\rangle=|3,3\rangle$ with impurity atoms of state $|F,m_F\rangle=|3,2\rangle$. The inset showed   scattering lengths between the host atoms $|3,3\rangle$-$|3,3\rangle$ and between host atom and impurity $|3,3\rangle$-$|3,2\rangle$. The host atoms were  levitated against the gravitational force, whereas  the gravitation force $F$ acted  on the impurity atoms. (B) The excitation spectrum of the impurity exhibited  cosine-shaped curve (solid black), whereas the particle-hole excitations of the host atoms had  a continuum spectra. When the impurity accelerated by the gravitational force reached  the edge of the Brillouin zone $k_F$, the Bragg reflection of the impurity atom occurred, see the inset, due to the absorption of the excitation with momentum $2k_F$ by the host atoms without energy cost.
  Figure from  \cite{Meinert:2017}. }
 \label{fig:Bloch-oscillaton}
 \end{figure}

The study of quantum physics of 1D integrable models are rapidly developing.
Recent developments of quantum solvable models comprise new frontiers in various fields such as  atomic physics, condensed matter physics, solid state materials and high energy physics, etc. Yang-Baxter integrability also lays out new trends in  quantum information, quantum metrology and quantum technology,
ranging from quantum heat engine \cite{Mukherjee:2021,Brantut:2013,Hofer:2017,Jaramillo:2016,Gluza:2021,ChenYY:2021}  and quantum refrigeration \cite{ Medley:2011,Weld:2010,Yu:2020} to quantum battery \cite{Strambini:2020,Caliga:2017}, atomtronics \cite{Amico:2021}, precision measurement \cite{Bentsen:2019,Cai:2021,Grun:2022,Kitagawa:2010},  quantum transport \cite{Bertini:2021}, quantum control \cite{Wilsmann:2018} and quantum impurities \cite{Mathy:2012,Meinert:2017,Knap:2014,Dolgirev:2021,Dehkharghani:2018}.
In this short review, our aim was not  to review all such developments.  Instead we briefly discussed some highlights:  several  Yang-Baxter solvable models, including  the Lieb-Liniger model \cite{Lieb-Liniger:1963}, the Yang-Gaudin model \cite{Yang:1967,Gaudin:1967}, the $SU(w)$ Fermi gases \cite{Sutherland:1968}, the GHD theory, the Haldane's FES and compared them with recent novel experiments.
In particular, the newly developed theory of GHD \cite{Castro:2016,Bertini:2016}  coming along with experimental observations of the  quantum Newtons's cradle and the corresponding description in terms of GHD has been discussed in details.
Several other important experiments on dynamical fermionization,  quantum criticality, Luttinger liquid, the Haldane's fractional exclusion statistics,  quantum holonomy, spin-charge separation and quantum impurity  have been discussed and highlighted.
It turns out that  integrable models of  ultracold atoms provide an  ideal platform to experimentally study quantum dynamics, hydrodynamics and thermodynamics at a many-body level. Here an outlook is made for future research  on  the Yang-Baxter integrability with ultracold atoms:

 {\bf (a) Sensoring gravitational force:} In a recent paper \cite{Meinert:2017}, the Bloch oscillation of an impurity atom was found in the continuum liquid of 1D interacting Bose gas, see Fig.~\ref{fig:Bloch-oscillaton} (A).
Adding an  impurity atom with the same mass $m$ into a degenerate 1D gas of Cesium atoms trapped in each  tube of the 3D ensemble, F. Meinert {\em et al.} observed Bloch oscillation in the dynamics of the impurity atom of hyperfine state $|F,m_F\rangle=|3,2\rangle$, on which  a (dimensionless) gravitation force ${\cal F}=F m/(\hbar^2n^3_{\rm 1D})$ acts,   here $n_{\rm 1D}$ was the density of the host atoms and $F$ denoted  the gravitational force.
It was  remarkable to find that the characteristic Bragg reflection occurred  in the absence of a lattice, which was considered  the key setting for  the usual Bloch oscillations caused by  periodic momentum dependent eigenstates.  That was  mainly because the magnon  excitation had a collective cosine-shaped dispersion which  resembled  the conventional dispersion of magnon in a 1D lattice, see Fig.~\ref{fig:Bloch-oscillaton} (B). The periodicity was 2$k_F$, where the Fermi momentum $k_F=n_{\rm 1D} \pi$. This gave  an effective Brillouin zone. The impurity atom thus exhibited  Bragg reflection at the edge of this Brillouin zone. The excess momentum dissolved into particle-hole excitations of the host atoms of the state $|F,m_F\rangle=|3,3\rangle$. Although the observed Bloch oscillations in such  interacting 1D Bose gas were not perfect due to the  strong many-body correlation effects, the experiment holds a promise for future quantum technology in sensoring weak force.

Nevertheless, nonequilibrium dynamics  of Newtons's cradle \cite{Kinoshita:2006}, hydrodynamics of quasi-1D trapped
gases in a double-well potential \cite{Schemmer:2019}, quantum walks and Bloch oscillations of bosons and fermions in a 1D lattice \cite{Preiss:2015} will lead to potential applications in high precision measurement of the gravitational force and testing the Einstein equivalence principle. These settings will highlight the importance of quantum coherence and entanglement in quantum many-body dynamics out-of-equilibrium, in particular in systems with higher mathematical symmetries.
Studying quantum transport and conductivity of 1D many-body systems in the frame work of the GHD and Bethe ansatz still remain an open challenge.

{\bf (b) Quantum many-body entanglement:}  The observation of spin-charge separation phenomenon \cite{Vijayan:2020,Senaratne:2021} open the possibility  to further study novel spin states and fractional quasiparticles in 1D ultracold atoms.
 The experiment \cite{Senaratne:2021} remarkably provides a conclusive observation of the spin-charge separation in Luttinger liquids and also goes beyond it. Further studies of spin coherent and incoherent Luttinger liquids in the 1D Hubbard model, quantum gases with $SU(2)$ and fermionic atoms with  higher spin symmetries etc. is highly desirable.
 On the other hand, it was recently shown \cite{Hauke:2016} that the dynamical response function of spinons can be used to  measure multipartite entanglement in quantum  spin systems. The key to this study relies on the connection between the quantum Fisher information and dynamic susceptibility. The former can be used as a witness of the multipartite entanglement in quantum many-body systems. This  was experimentally demonstrated \cite{Scheie:2021,Laurell:2022,Zhang:2014} that the neutron scattering DSF can witness the entanglement of spins in the  1D Heisenberg chain. This opens promising opportunities to explore realistic applications of fractional excitations, spin liquids and impurities in quantum metrology.  Moreover, quantum entanglement in  1D  topological matter \cite{Leseleuc:2019,Kanungo:2022,Atala:2022} and integrable models with Lindblad dissipations  \cite{Ziolkowska:2020,Bacsi:2020,Nakagawa:2021,Landi:2022,Popkov:2022,Rosso:2022} comprise new frontiers in quantum integrability.

{\bf (c) Quantum heat engine and refrigeration:} Many-body physics presents fundamental principles which will play a key  role  in  future quantum technologies. The study of quantum cooling and heat engine \cite{Mukherjee:2021,Brantut:2013,Hofer:2017,Jaramillo:2016,Gluza:2021,ChenYY:2021} in ultracold atomic physics still remains rather elusive. To achieve the goal of cooling interacting fermions, it requires us to understand caloric effects induced by magnetic field, trapping potential and dynamical interaction as well as adiabatic evolutions of the energy and currents during strokes \cite{ Medley:2011,Weld:2010,Yu:2020,Daniloff:2021}.
For those  research, one has  many open questions regarding the quantum speed of  adiabatic processes and heat exchanges between the system and baths and caloric effects.
 The interaction ramp up and down near a quantum phase transition in quantum gases provide a promising protocol of quantum refrigeration besides  the usual adiabatic demagnetization cooling in solid-state materials \cite{Wolf:2011}.
Exactly solvable models of quantum gases, strongly correlated electrons and spins thus exhibit  rich phases of quantum matter which will provide important benchmarks and quantum advantage for studying  quantum heat engine, quantum refrigeration  and quantum batteries in quantum technology. And more  generally,  mathematical models   provide an ideal platform  to advance our  understanding of new quantum effects for future technology,  including quantum metrology, quantum information and quantum communication.

\acknowledgments

X.W.G. is supported by the NSFC key grant No. 12134015, the NSFC grant  No. 11874393 and No. 12121004.
The authors thank Angela Foerster, H Pu, Hui Hu, Xia-Ji Liu,  Zhen-Sheng Yuan, Shi-Guo Peng, Sheng Wang for helpful discussions.
They thank  Randy Hulet and Natan Andrei for their  help with going through the manuscript.

\end{document}